\documentclass[onecolumn, 11pt, 3p]{elsarticle}

\usepackage{hyperref}
\usepackage[normalem]{ulem}
\usepackage{xcolor} 
\usepackage[nomessages]{fp}
\usepackage{booktabs}
\usepackage{multirow}
\usepackage{url}
\usepackage{framed}
\usepackage{enumitem}
\usepackage{array}
\usepackage{tabularx}
\usepackage{balance}
\usepackage{amsmath}
\usepackage{siunitx}
\usepackage{subcaption}
\usepackage{booktabs}
\usepackage{xspace}
\usepackage{adjustbox}
\usepackage[flushleft]{threeparttable}
\usepackage{setspace}
\usepackage{caption}
\usepackage[font=itshape]{quoting}
\usepackage{arydshln}
\usepackage{soul,color}

\usepackage{tcolorbox}
\tcbuselibrary{breakable}

\soulregister\cite7
\soulregister\ref7
\soulregister\pageref7
\setstcolor{red}
\setul{}{.2ex}

\newcommand{\politifact}{\texttt{PolitiFact}\xspace}
\newcommand{\abc}{\texttt{ABC}\xspace}
\newcommand{\politifactpantsfire}{\texttt{pants-on-fire}\xspace}
\newcommand{\politifactfalse}{\texttt{false}\xspace}
\newcommand{\politifactbarelytrue}{\texttt{barely-true}\xspace}
\newcommand{\politifacthalftrue}{\texttt{half-true}\xspace}
\newcommand{\politifactmostlytrue}{\texttt{mostly-true}\xspace}
\newcommand{\politifacttrue}{\texttt{true}\xspace}
\newcommand{\abcpositive}{\texttt{positive}\xspace}
\newcommand{\abcnegative}{\texttt{negative}\xspace}
\newcommand{\abcinbetween}{\texttt{in-between}\xspace}

\newcommand{\completelydisagree}{\texttt{Completely Disagree (-2)}\xspace}
\newcommand{\disagree}{\texttt{Disagree (-1)}\xspace}
\newcommand{\neitheraord}{\texttt{Neither Agree Nor Disagree (0)}\xspace}
\newcommand{\agree}{\texttt{Agree (+1)}\xspace}
\newcommand{\completelyagree}{\texttt{Completely Agree (+2)}\xspace}

\journal{Journal of Information Processing and Management}

\biboptions{numbers}

\usepackage{array}
\newcolumntype{P}[1]{>{\raggedright\arraybackslash}p{#1}}

\begin{document}

\begin{frontmatter}
\date{July 28, 2021}
\title{The Many Dimensions of Truthfulness:\\ Crowdsourcing Misinformation Assessments on a Multidimensional Scale}

\author[1]{Michael Soprano}
\author[1]{Kevin Roitero}
\author[1]{David La Barbera}
\author[2]{Davide Ceolin}
\author[3]{Damiano Spina}
\author[1]{Stefano Mizzaro}
\author[4]{Gianluca Demartini}

\address[1]{University of Udine, Udine, Italy.}
\address[2]{Centrum Wiskunde \& Informatica (CWI), Amsterdam, The Netherlands.}
\address[3]{RMIT University, Melbourne, Australia.}
\address[4]{The University of Queensland, Brisbane, Australia.}

\begin{abstract}
Recent work has demonstrated the viability of using crowdsourcing as a tool for evaluating the truthfulness of public statements. Under certain conditions such as: (1) having a balanced set of workers with different backgrounds and cognitive abilities; (2) using an adequate set of mechanisms to control the quality of the collected data; and (3) using a coarse grained assessment scale, the crowd can provide reliable identification of fake news. However, fake news are a subtle matter: statements can be just biased (``cherrypicked''), imprecise, wrong, etc. and the unidimensional truth scale used in existing work cannot account for such differences. 
In this paper we propose a \textit{multidimensional} notion of truthfulness and we ask the crowd workers to assess seven different dimensions of truthfulness selected based on existing literature: Correctness, Neutrality, Comprehensibility, Precision, Completeness, Speaker’s Trustworthiness, and Informativeness. We deploy a set of quality control mechanisms to ensure that the thousands of assessments collected on 180 publicly available fact-checked statements distributed over two datasets are of adequate quality, including a custom search engine used by the crowd workers to find web pages supporting their truthfulness assessments.
A comprehensive analysis of crowdsourced judgments shows that:
(1) the crowdsourced assessments are reliable when compared to an expert-provided gold standard; (2) the proposed dimensions of truthfulness capture independent pieces of information; (3) the crowdsourcing task can be easily learned by the workers; and (4) the resulting assessments provide a useful basis for a more complete estimation of statement truthfulness.
\end{abstract}

\begin{keyword}
Truthfulness, Crowdsourcing, Misinformation, Explainability
\end{keyword}

\end{frontmatter}

%\linenumbers

\section{Introduction}
The spread of online misinformation  has important effects on the stability of the democratic process and on human decision making processes \cite{visser2020reason}.
The sheer size of digital content in the web and social media and the ability to immediately access  and share it have made it difficult to perform timely fact-checking at scale.
While significant efforts have been made by different research communities to automatize fact-checking (see, e.g., \cite{atanasova2019automatic,elsayed2019overview,hansen2019neural}), it is often still necessary to involve humans in the fact-checking process. While leveraging experts to judge and render a verdict on the truthfulness of news is the common approach, this becomes too expensive and impractical if %required to be 
performed at scale.
Thus, recent work has looked at the possibility to employ crowdsourcing methods to perform fact-checking at scale \cite{la2020crowdsourcing,roitero2020crowd,roitero2020covid}. While truthfulness scales at different levels of granularity have been compared leading to the conclusion that coarse-grained (e.g., three levels) scales are to be preferred for crowdsourced truthfulness annotations \cite{la2020crowdsourcing}, a uni-dimensional truthfulness scale  appears to be too simplistic to  capture all the nuances of truthfulness.

In this paper, we study how crowdsourcing truthfulness annotation tasks may be performed by taking a \textit{multidimensional} labeling approach rather than asking annotators to label on a single scale between the `true' and `false' extremes.
Specifically, we deployed a task asking US-based crowd workers recruited from Amazon MTurk to label the truthfulness of political statements not just based on a single multi-level scale (e.g., like done by \citet{politifact} with a 6-level scale), but rather using multiple dimensions of truthfulness. We asked participants to label a statement on a scale for each of the Correctness, Neutrality, Comprehensibility, Precision, Completeness, Speaker’s Trustworthiness, and Informativeness dimensions.

Our results show that:
the truthfulness judgments provided by crowd workers over the different dimensions are sound, reliable, and independent; 
the agreement between crowd and expert judgments is good for the Overall Truthfulness;
the crowd labels are informative about the reasons underlying  crowd judgments and are difficult to be generated automatically;
finally, we  show that implicit signals from crowd workers can be leveraged to effectively predict the expert judgments, across all the datasets we considered.

This paper is structured as follows. In Section \ref{sect:related} we survey  related work. In Section \ref{sect:aims} we detail our research questions, addressed using the experimental setting described in Section \ref{sect:experiment}. In Section \ref{sect:results} we analyze  the results. In Section \ref{sect:discussion} we discuss our findings and conclude the paper.

\section{Related Work}
\label{sect:related}

We survey in the following subsections the background work on different aspects of misinformation:
the use of automated algorithms (Section~\ref{sect:bg-automated-fact-checking}),
the use of crowdsourcing (Section~\ref{sect:bg-crowdsourcing-truthfulness}),
the phenomena of echo chambers and filter bubbles (Section~\ref{sect:bg-echo-chambers}),
the application of argumentation theory (Section~\ref{sect:bg-argument-mining}), and 
the use of multidimensional scales (Section~\ref{sect:bg-multidimensional}).

\subsection{Automated Fact-Checking} \label{sect:bg-automated-fact-checking}

Automated fact-checking aims at replacing experts, i.e., usually journalists, in performing this task.
As an example of such methods, \citet{10.1145/3386253} proposed a deep neural network model to detect misinformation statements. Their model is based on a feature extractor which works both at the textual and at the user level, an attention layer used to detect important and specific user responses, and a pooling algorithm to do feature aggregation. Their results on two datasets show that the developed model reaches an accuracy level higher than $0.9$ within 5 minutes from the spread of the misinformation statement.
As for another example, \citet{limetal2020annotating} used crowdsourcing to gather bias labels on news articles and proposed an automatic approach for analyzing and detecting it. \citet{li2020misinformation}, instead, proposed to identify possible misinformation on Twitter by learning a topic-based model from expert provided assessment. 
However, as evidenced by the approaches that exploit machine learning to build completely automatic classifiers, fact-checking still requires manual effort, usually from expert fact-checkers, to generate labels that can eventually lead to the training of supervised methods like the one described above. So, the work presented in this paper is complementary to this.

\subsection{Crowdsourcing Truthfulness}  \label{sect:bg-crowdsourcing-truthfulness}
Previous work has looked at how to use crowdsourcing to collect truthfulness labels in order to scale-up the manual fact-checking effort \cite{visser2020reason}.
\citet{la2020crowdsourcing} extended a work by \citet{roitero2018many} by assessing the truthfulness of political statements by means of crowdsourcing and focusing on the effects of the assessor bias and judgment scale. Their results show that workers have a preference toward scales with less values (i.e., coarse-grained), and that there is a strong effect of the assessor political background on their ability to effectively assess misinformation statements.
\citet{roitero2020crowd} investigated whether the crowd can objectively identify and classify misinformation statements. To this aim, they collected thousands (i.e., 5,400, or 6,600 including the gold questions) assessments for political statements taken from two popular datasets used to perform fact-checking using three scales. Their results show that grouping adjacent ground truth categories together lead to a high agreement between crowd and expert labels, indicating that workers are able to distinguish between true and false statements. They also found that the different scales lead to similar agreement levels, and that the workers bias has an impact on assessment quality.
\citet{roitero2020covid} followed the same approach of \citet{roitero2020crowd} to study if the crowd can reliably assess misinformation statements related to the COVID-19 pandemic. Their results, apart from reporting the effectiveness of crowd workers, study many aspects such as agreement, workers background, bias, and behavior.
Related to these works, \citet{10.1145/3313831.3376232} conducted a survey experiment with about 1,000 Americans to understand their perceived trust in numerous news sites; their results show that participants tend to trust mainstream sources more than hyper-partisan or fake news sources. \citet{bhuiyan2020investigating}, instead, collected credibility annotations on the topic of climate change from both crowd workers and students with journalism or media programs; they studied and compared the two sets of annotations against expert-provided ones.
\citet{10.1145/3340531.3412169} introduced a tutorial on online harmful information that includes social media and fake news.
As compared to this body of work, in our paper we investigate the effect of asking crowd assessors to judge truthfulness using multiple dimensions and observe if doing so has an impact on the quality of the collected labels.

\subsection{Echo Chambers and Filter Bubbles}  \label{sect:bg-echo-chambers}
Related works also addressed the way information spreads through social media and, in general, the Web, leading to the discovery of a number of phenomena that were not so evident before. Among those, \emph{echo chambers} and \emph{epistemic bubbles} seem to be central concepts \cite{nguyen_2020,pariser2011filter}. \citet{doi:10.1177/2158244019832705} investigated the extent of ideological echo chambers on social media using well-known media organisations and political actors as anchors. \citet{flaxman2016filter} mined search history of U.S. users to investigate the effect of search engines and social networks in the user's opinions and exposure to news.%

\subsection{Truthfulness and Argument Mining} \label{sect:bg-argument-mining}

Truthfulness classification and the process of fact-checking are strongly related to the scrutiny of factual information  extensively  studied  in  argumentation  theory \cite{lawrence2020argument,williams2001application}. 
Argument mining, i.e., the automatic identification and extraction of the structure of inference and reasoning expressed as arguments presented in natural language, is also related to our work. \citet{lawrence2020argument} surveyed the techniques used for argument mining and detailed how crowdsourcing based approaches can be used to overcome the limitations of manual analysis. \citet{sethi2017spotting} proposed a prototype social argumentation framework to curb the propagation of fake news where the argumentation structure is crowdsourced and reviewed/moderated by a set of experts in a virtual community. \citet{visser2020reason} showed how to use argument mining to increase skills of workers that assess media reports. \citet{sethi2019fact} developed a recommender system that makes use of argumentation and pedagogical agents in order to fight misinformation. 
Such argumentation frameworks can be used to leverage quality of crowdsourced items, e.g., by providing to the crowd workers some tools to better assess the argument structure of statements. Also, 
\citet{DBLP:conf/icwe/CeolinPWS21} explore the relation between multidimensional information quality assessment and argumentation reasoning, highlighting the fact that argumentation reasoning can identify items showing particular aspects of quality (e.g., accuracy, readability) in the case studies addressed.

\subsection{Multidimensionality of Relevance and Truthfulness Judgments} \label{sect:bg-multidimensional}

Related work has also looked at how human assessors perform judgments when using multiple dimensions and at comparing experts and non-experts. 
Multidimensional scales proved to be effective in the setting of information retrieval when dealing with relevance.
\citet{BARRY1998219}, in their classical work, and \citet{10.5555/1133031.1133039} listed the different relevance criteria used to perform relevance evaluation;
\citet{jiang2017comparing} collected multidimensional relevance along with contextual feedback from users and correlate their judgments with user metrics.
\citet{uprety2020quantum} defined multidimensional relevance using a quantum inspired structure. %
\citet{10.1145/2600428.2609577} 
extended the psychometric framework for multidimensional relevance proposed by \citet{10.1007/s10791-012-9206-z} by using crowdsourcing, detailing its limitations, and describing various quality control methods derived from psychometric which can be applied to the information retrieval context.
\citet{jiang2017comparing} investigated two variants on TREC-style relevance judgments used in information retrieval: they studied contextual judgments and they collected multidimensional judgments, using novelty, understandability, reliability, and effort as dimensions. 

Given the amount of research done and the demonstrated effectiveness of multidimensional relevance judgments, it seems natural to try and apply the same approach to truthfulness judgments. There is indeed some preliminary work in this direction.
\citet{ceolin2016capturing} collected multidimensional truthfulness judgments on web documents dealing with vaccines, where few experts provided the assessments.
Their results showed that experts manifest a high level of agreement, but also that the task is very demanding, and that the availability of experts online is rather limited.
\citet{maddalena2018multidimensional} extended the work by \citet{ceolin2016capturing} by comparing crowd and expert truthfulness assessment for a small dataset of 20 selected documents dealing with vaccines. Results show that experts inclined to use lower values than crowd workers (i.e., they are more critical), and that the agreement between crowd and experts is high, but not total.

To the best of our knowledge, our paper is the first aiming at collecting a large amount of truthfulness judgments on a multidimensional scale, and making it available to the research community.

\section{Aims and Research Questions}
\label{sect:aims}

We ran a large scale crowdsourcing experiment and asked crowd workers to assess political statements with the aim of identifying online misinformation. We used the same set of statements used by \citet{roitero2020crowd}, which is publicly available and has been fact-checked by experts. Differently from previous work, we used a multidimensional notion of truthfulness, detailed in Section~\ref{sec:7dim}, and collected independent judgments for each dimension from each worker. The workers also assessed the Overall Truthfulness of each statement and they had to justify their choice by providing a URL to the web page they used to verify the truthfulness of the statement.

We focused on the following specific research questions: 
\begin{enumerate}[label={RQ\arabic*}]

     \item\label{i:crowd-reliability} Are crowd workers able to reliably assess multiple dimensions of information truthfulness? How do their judgments correlate with expert judgments?
     
     \item\label{i:rrq2}
     Are all truthfulness dimensions independent from each other, and thus required? Can some dimensions be derived from (a combination of) the others?
     Is it possible to combine the individual dimensions in a way that it improves agreement between crowd and expert judgments?
 
    \item\label{i:worker-behavior} What is the behavior of workers when choosing labels for truthfulness dimensions? Do their cognitive abilities have any influence? 
     
     \item\label{i:dim-informativeness} How meaningful and informative are the individual information quality dimensions?
     
     \item\label{i:machine-learning} Can the multidimensional judgments be used to accurately predict the expert judgments and verdicts?

\end{enumerate}

\section{Experimental Setting}
\label{sect:experiment}

In this section we outline the composition of the dataset used in this paper (Section~\ref{sect:dataset}), the dimensions of truthfulness that we used in our experiment (Section~\ref{sec:7dim}), and the design of the crowdsourcing task (Section~\ref{subsect:task}) used to collect the set of judgments. Section~\ref{sect:desc} reports some descriptive statistics about the crowd workers and the collected judgments, and Section~\ref{sect:abandonment} analyzes the behavior of the workers that abandoned the task without completing it.

\subsection{Dataset}
\label{sect:dataset}

We used 180 political statements sampled from two different datasets (i.e., collections of statements), namely \politifact and \abc. 

\politifact \cite{politifact} is a publicly available dataset dedicated to fake news detection and contains more than 12,800 human labeled short statements. The speakers in such a dataset include members of U.S. political parties, as well as a significant amount of posts from online social media. 
Human editors evaluated each statement using a six-level truthfulness scale: \politifactpantsfire, \politifactfalse, \politifactbarelytrue, \politifacthalftrue, \politifactmostlytrue, and \politifacttrue.

\begin{table*}[tb]
\centering
\caption{Example of \politifact (first row) and \abc (second row) statements. \abc statement shows both simplified truthfulness judgment and original verdicts.
}
\label{tab:statements}
{\small
\begin{tabular}{@{}p{6cm}@{\quad}p{1cm}@{\quad}l@{\quad}l@{}}
\toprule
  \textbf{Statement} & \textbf{Source} & \textbf{Year} & \textbf{judgment / Verdict}\\
 \midrule
  ``Washing your hands and covering your mouth when you cough makes a huge difference in reducing transmission of the flu.'' & Barack Obama & 2009 & \mbox{\politifacttrue} \\ 
  \midrule
  ``Under this government, the tax to GDP ratio has, in the period weve been in office, [been] an average of 22.7 per cent'' & Kevin Rudd & 2013 & \mbox{\abcpositive} /  \texttt{Checks Out} \\ 
\bottomrule
\end{tabular}%
}
%\end{adjustbox}
\end{table*}

\abc\footnote{\url{https://apo.org.au/collection/302996/rmit-abc-fact-check}} is a collection of about 500 statements (as of today) claimed between 2013 and 2020 and verified by Australian Broadcasting Corporation which aims to determine the accuracy of claims by Australian politicians, public figures, advocacy groups, and institutions engaged in public debate. When there is a statement to check, a researcher contacts various experts in the field to seek their opinion and guidance on the available evidence. Then, the chief fact checker reviews the statement and after such step the team discusses the final verdict. The available verdicts are heterogeneous---up to 30 different verdicts in the sample we used in our experiments---but there is also a simplified version of such verdict provided using a three-level scale: \abcnegative, \abcinbetween, and \abcpositive. 

To select the statements to be assessed, we relied on the choice made by \citet{roitero2020crowd}. Thus, we selected the same statements as they did, to directly see the impact of a multidimensional scale, as well as to provide the research community with two sets of annotations referring to the same set of statements.
\citeauthor{roitero2020crowd} selected 10 statements for each of the two political parties, for each truthfulness level. The \politifact dataset contains statements given by U.S. politicians (Democratic and Republican parties) using a six-level truthfulness scale; this means that a total of $10 * 2 * 6 = 120$ statements were sampled. The \abc dataset concerns statements given by Australian politicians (Labor and Liberal parties) using a three-level truthfulness scale; this means that a total of $10 * 2 * 3 = 60$ statements were sampled. Therefore, the total amount of sampled statements is 180. Table~\ref{tab:statements} shows a sample of the statements that we used.%

\subsection{The Seven Dimensions of Truthfulness}
\label{sec:7dim}

The main difference of our experimental setting from the one employed by \citet{roitero2020crowd} is that each worker was asked to assess seven different dimensions of truthfulness more than just the Overall Truthfulness of the statement. We chose to use the following dimensions reported here as presented to the workers, who were also shown an example for each dimension. A detailed description of each dimension and the examples provided to the workers can be found in \ref{app:instructions}.%

\begin{enumerate}[]
\item \textit{Correctness}: the statement is expressed in an accurate way, as opposed to being incorrect and/or reporting  mistaken information.
\item \textit{Neutrality}: the statement is expressed in a neutral / objective way, as opposed to subjective / biased. 
\item \textit{Comprehensibility}: the statement is comprehensible / understandable / readable as opposed to difficult to understand. 
\item \textit{Precision}: the information provided in the statement is precise / specific, as opposed to vague.
\item \textit{Completeness}: the information reported in the statement is complete as opposed to telling only a part of the story.
\item \textit{Speaker’s Trustworthiness}: The speaker is generally trustworthy / reliable as opposed to untrustworthy / unreliable / malicious.
\item \textit{Informativeness}: The statement allows us to derive useful information as opposed to simply stating well known facts and/or tautologies.
\end{enumerate}

The choice of dimensions is informed by previous work.
In the information systems literature, information quality and user satisfaction are
two major dimensions for evaluating the success of information systems \cite{10.1145/505248.506007}. These two facets can be further split along different characteristics. Given that we are mainly interested in news truthfulness, we focused on information quality characteristics, such as accuracy and precision. 
The ISO 25012 Model derived these dimensions from various related works \cite{10.1145/505248.506007, wang1996beyond, vanivcek2005software}. 
The dimensions of Correctness, Completeness, Precision, Comprehensibility, and Neutrality considered in our work are thus motivated by the ISO Model \cite{ISOmodel} and are intended to describe information quality. In addition, we also considered two additional dimensions, Speaker's trustworthiness and Informativeness, which find motivations in the literature; \citet{jowett2018propaganda} highlighted the influence of the speaker’s trustworthiness in relation to the judgment towards a statement the reliability of the source is one of the relevance dimension catalogued in the work by \citet{BARRY1998219}. \citet{maddalena2018multidimensional} and \citet{ceolin2016capturing} used Informativeness among other dimensions to perform crowdsourcing tasks dealing with information quality assessment. It is important to note that these additions are necessary, since the ISO model focuses on data quality, while here we are interested in assessing the quality of the information represented by such data. Thus, we considered the subset of dimensions from the ISO model that are relevant in this context, and we extend them with additional ones motivated by the literature.

In more detail, in our work we considered the same dimensions employed by \citet{maddalena2018multidimensional}, who performed a crowdsourcing experiment with the aim of understanding if the crowd is a valid alternative to the experts for the task of information quality assessment. \citeauthor{maddalena2018multidimensional} used almost the same dimensions previously detailed by \citet{ceolin2016capturing}, who presented an experiment aimed to perform user studies considering web documents about the vaccination debate. \citet{maddalena2018multidimensional} slightly reformulated the description of some dimensions to adapt them and make them more adequate for the crowdsourcing context. Both studies found that using such a set of dimensions, crowd workers and experts perform well reaching a satisfactory level of external agreement when comparing the crowd and expert labels. 
Summarizing, we chose to consider those particular seven dimensions because they find a theoretical grounding and are proven to lead to a good level of external agreement, allowing us to capture information accuracy and appropriateness.%

\subsection{Crowdsourcing Task}
\label{subsect:task}

We used the Amazon Mechanical Turk\footnote{\url{https://www.mturk.com/}} crowdsourcing platform to collect data. When a worker accepted a Human Intelligence Task (HIT), he/she was shown an input token and a URL to an external server which contained a deployment of our web application (i.e., the actual task). The worker carried out the assigned HIT on such an application. If s/he successfully completed the HIT he/she was shown an output token, which had to be copied back to the MTurk page to receive the payment upon approval. 

Each HIT of our crowdsourcing task followed a design similar to that used by \citet{roitero2020crowd}. Each crowd worker is firstly asked to fill a mandatory questionnaire composed of seven questions to collect his/her background information. 

We asked them about their age, instruction level, and family income. Then, we turned to their political views and we asked how he/she identified such views and in which political party. As for the last two question, we asked workers's opinion on climate change and U.S. southern border. After this first questionnaire, the worker was asked to answer three modified Cognitive Reflection Test (CRT) questions, originally proposed by \citet{Frederick2005}, which are used to measure whether a person tends to overturn the incorrect “intuitive” response and further reflect based on their own cognition to find the correct answer. Such CRT questions allow to assess the cognitive abilities of a person. 

After the questionnaires, the worker was asked to assess 11 statements selected from \politifact (6 statements) and \abc (3 statements) dataset.
Each HIT contained a statement for each truthfulness label of the \politifact and \abc datasets, plus 2 special statements used for the purpose of quality checks. 
We built each HIT using a randomization process to avoid all the possible source of bias.

In more detail, the crowd worker was first asked to provide the Overall Truthfulness of the statement and a Confidence level of his/her knowledge of the topic. Then, the worker had to provide the URL that he/she used as a source of information to assess the Overall Truthfulness. Such a URL had to be found using a customized search engine (implemented using Microsoft Bing Web Search API\footnote{\url{https://www.microsoft.com/en-us/bing/apis/bing-web-search-api}} and available to the workers right below the statement) which allows to filter out \politifact and ABC websites from search results. To ensure that the workers do not bypass our search engine, we also checked if the selected URL was one of the ones retrieved by our own search engine, otherwise the user was not allowed to proceed in the task. 
Then, each worker was also asked to assess the seven different dimensions of truthfulness described in Section~\ref{sec:7dim}. Each judgments was expressed on the following Likert scale \cite{likert}: \completelydisagree, \disagree, \neitheraord, \agree, \completelyagree. 
The set of instructions shown to the workers and containing a detailed description of the assessment process is available in \ref{app:instructions}.

Besides the above described controls on the URL, we also implemented different quality checks to ensure the high quality of the collected data: two gold questions (i.e., two statements which are clearly true or false, and checking the consistency of the answers), 
monitoring of the time spent (i.e., checking whether the workers spent at least 2 seconds on each statement).

Overall, we used 180 statements in total as outlined in Section~\ref{sect:dataset}, and each statement was evaluated by 10 distinct crowd workers. Thus, we deployed 200 MTurk HITs and we collected 2200 judgments in total. Each worker reward was of  2\$ for a task including the set of 11 judgments, computed on the basis of the time needed to finish the task and the U.S. Minimum Salary Wage of 7.25 USD per hour. 
The crowd task was launched on June 1st, 2020 and it finished on June 4th, 2020.

The dataset used to carry out our experiments can be downloaded at \url{https://github.com/KevinRoitero/crowdsourcingTruthfulness}. 
This study has been approved by The University of Queensland Ethics Committee.
Participating workers were presented with an information sheet that detailed the worker's rights, which data would have been collected, how the data would have been used, the outcome of the project, and the compliance and consistency with data protection laws. We made sure that participation was anonymous, and that any data related to topics of sensitive nature was not collected or stored. Also, participation did not involve psychological distressing search tasks. All collected data was securely stored on our database.
If workers wish at any time to withdraw their participation after submitting their answers, they can request for the collected data to be deleted by sending an email to any of the team members listed in the information sheet.

\subsection{Descriptive Statistics}\label{sect:desc}

Overall, 200 crowd workers successfully completed the experiment.
Amazon MTurk allows to select workers living within a certain country and each worker must provide some personal info when subscribing such as the home address. We have requested only U.S. citizens and we derived the following demographic statistics. Nearly  49\% of workers (95/200) are between 26 and 35 years old. The majority of workers (52\%) have a college/bachelor degree. As for total income before taxes, the 22\% earned 50,000\$ to less than 75k\$, while the 18\% earned 40k\$ to less than 50k\$. Turning to their political views, the 33\% identified their political views as Liberal, the 22\% as Moderate, and the 16\% as Conservative. The majority of workers (46\%) considered themselves as Democrats, while the 28\% as Republicans and the 23\% as Independent. The majority of workers (53\%) disagreed with building a wall on U.S. southern border while the 40\% agreed.
Finally, the vast majority of workers (85\%) thought that the government should increase environmental regulations to prevent climate change, while only the 9\% disagreed.
In general, we can say that our sample is well balanced, with the only exception of a few categories. Overall, our sample is in line with previous studies.

\subsection{Task Abandonment}\label{sect:abandonment}

To quantify how many workers abandoned the task we measured the abandonment rate using the definition provided by \citet{8873609}. Overall, we found that 200/681 workers (about 29\%) successfully completed the task while 355/681 workers (about 52\%) abandoned it (i.e., voluntarily terminated the task before completing it), and 126/681 (about 18\%) failed (i.e., terminated the task due to failing the quality checks too many times). Furthermore, 
184/651 workers (about 27\%) abandoned the task before really starting it; in other words, right after the completion of the initial questionnaire.%

 \begin{figure}[tbp]
   \centering
   \begin{tabular}{@{}c@{}c@{}}
   \includegraphics[width=.48\linewidth]{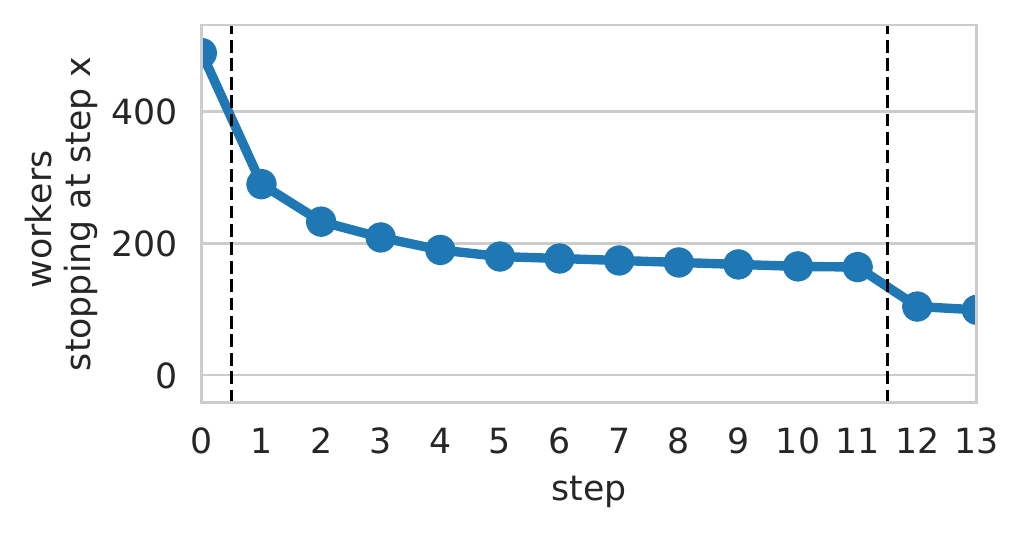}&
     \includegraphics[width=.48\linewidth]{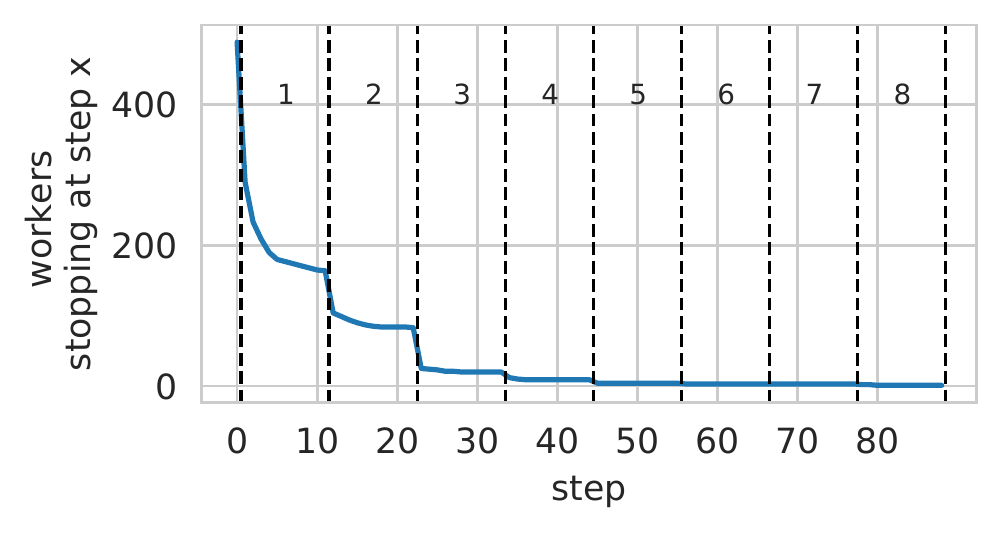}
   \end{tabular}
 \caption{
Abandonment rate shown as number of workers that reached a certain number of steps in the task: with a focus on the questionnaire and the first attempt (left plot) and over all attempts (right plot).
The abandonment monotonically decreases as the step number increases.
}
   \label{fig:abandonment}
 \end{figure}

 Figure~\ref{fig:abandonment} left plot shows the abandonment rate breakdown across task steps. A worker reached the next step when he/she completed the assessment of a single statement. Therefore, a task is completed if the worker assessed each statement within his/her current attempt. It must be noted that this definition does not make any assumption on task success. 
Step $0$ is the questionnaire, and each submit attempt occurred every 11 steps (since each HIT is composed by 11 statements). 
 As it can be seen, the abandonment rate monotonically decreases when the step number increases. There are two consistent drops of such amount that occur (highlighted by the dashed vertical lines in figure). Many workers abandoned the task when they completed only the questionnaires, i.e., at step $0$. The second drop occurred between step $11$ and step $12$, i.e., when they completed and failed the first attempt thus becoming bored or frustrated. Some workers performed up to 8 attempts before abandoning the task. These abandonment distributions are aligned with those found in previous work \cite{8873609, roitero2020crowd, roitero2020covid} and thus provide a first confirmation of the quality of the data.

The workers could leave an optional comment in a text field at the end of the task. When analyzing those comments, we observed that while some workers expressed minor concerns about the quality of the results  returned by our search engine or other minor issues, the vast majority of workers stated that they enjoyed the task and asked us to provide them with other similar tasks. Such comments and the analyses on abandonment rate provide a first indication that the task was performed accurately by the workers.

\section{Results}
\label{sect:results}

We report our results by addressing and discussing our five research questions in each of the following subsections.

\subsection{\ref{i:crowd-reliability}: Reliability of Multidimensional Assessment}
\label{sect:crow-accuracy}

We address the reliability of multidimensional assessment by analyzing: (i) the distributions of the individual and aggregated judgments provided by crowd workers; (ii) the internal agreement (i.e., the agreement measured among workers); (iii) how their judgments correlate with the judgments provided by experts; and (iv) their behavior while assessing each truthfulness dimension.

\subsubsection{Distributions of Judgments}

\begin{figure*}[htp!]
  \centering
  \begin{tabular}{@{}c@{}c@{}}
   \includegraphics[width=1\textwidth]{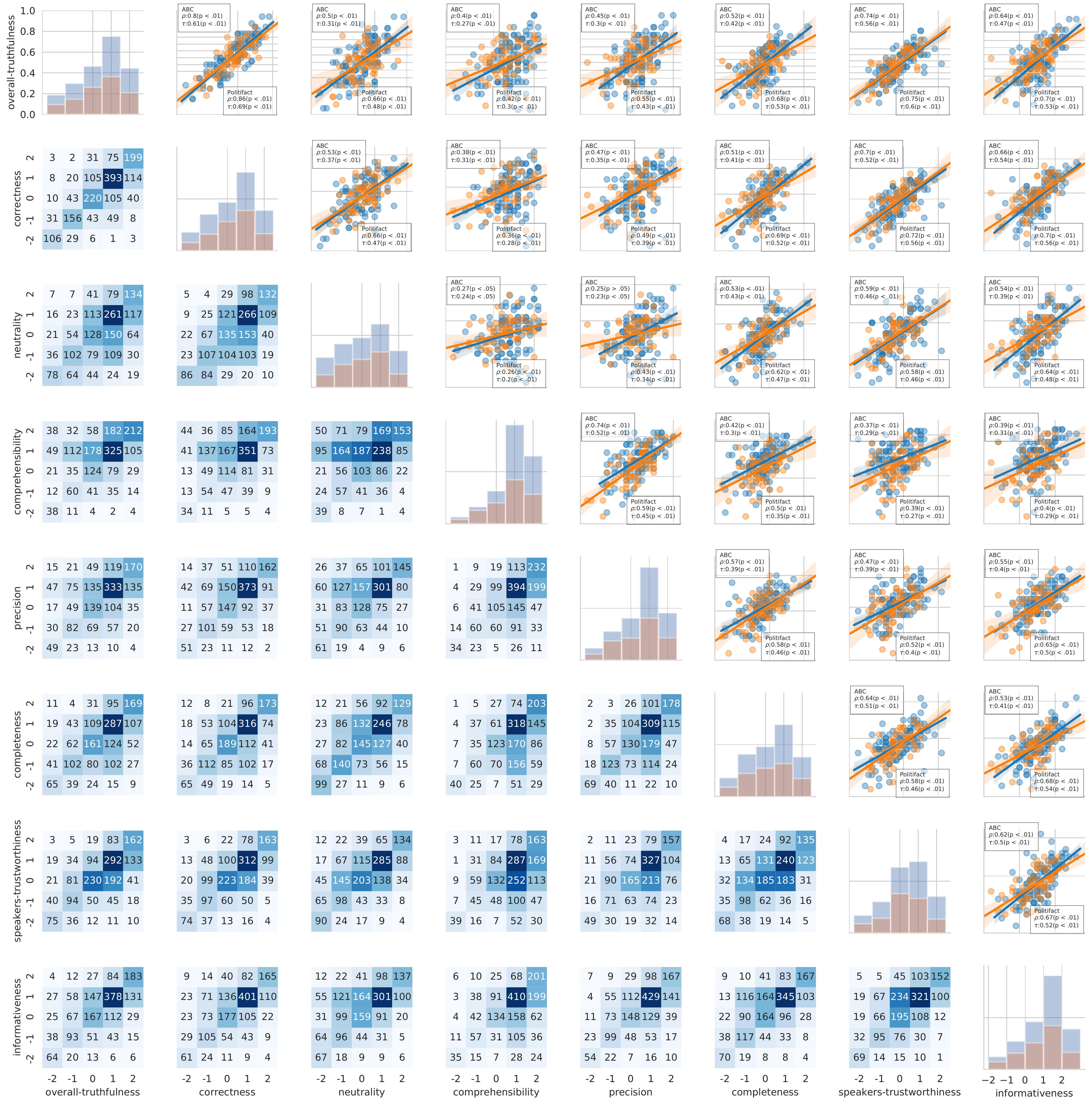}\\
   \includegraphics[width=.99\linewidth]{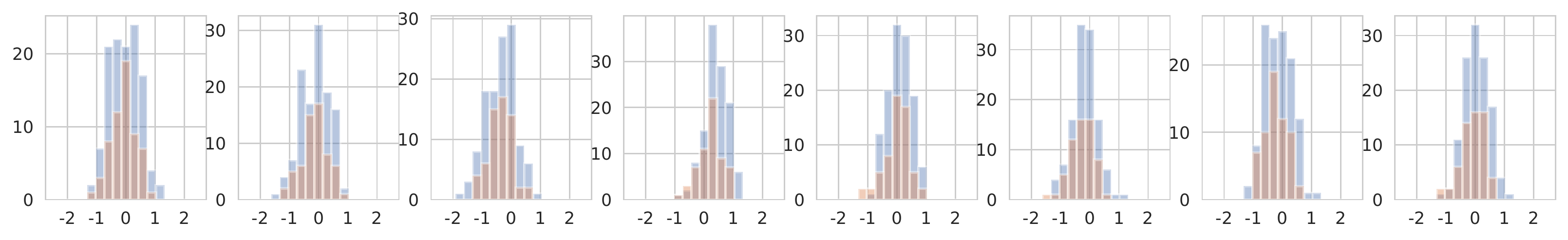}
  \end{tabular}
\caption{%
Correlation between dimensions: 
individual in the lower triangle and  diagonal,
aggregated in the upper triangle, aggregated distribution in the last row; breakdown on  \politifact (in blue) and \abc (in orange) categories (better on screen and using the zoom feature). 
Workers values are skewed towards positive values, i.e., \agree and \completelyagree (diagonal and bottom plots), and different dimensions have different correlation values (upper and lower triangle).}
  \label{fig:correlation-between-dimensions}
\end{figure*}

We start by analyzing Figure~\ref{fig:correlation-between-dimensions}. The heatmaps in the lower triangular matrix show the individual judgments collected for each dimension. There is a total of 28 heatmaps, one for each pair of dimensions. For each heatmap, each cell shows how many times the judgments are equal for the considered pair of dimensions. The histograms on the diagonal of Figure~\ref{fig:correlation-between-dimensions} show the distributions of the individual judgments for both \politifact (blue) and \abc (orange), for each dimension. Note that we collected half of \abc judgments compared to \politifact. We can see that each distribution is skewed to the right (i.e., towards higher truthfulness values) showing that workers tend to agree with statements more than disagree, or at least to not have a strong opinion. Since our subset of statements is balanced, as described in Section~\ref{sect:dataset}, this means that workers tend to agree also with false statements. However, this may be due to the scale used, which is different with respect to the original \cite{roitero2020crowd}.
The individual judgments are then aggregated using arithmetic mean since previous work \cite{roitero2018many,la2020crowdsourcing,roitero2020crowd} shows that it allows to obtain better result. The scatterplots in the upper triangular matrix show how the aggregated judgments of each pair of dimensions correlate, for both \politifact (blue) and \abc (orange). Each point within a plot represents a statement. The histograms on the bottom row of Figure~\ref{fig:correlation-between-dimensions} show the distributions of the aggregated judgments for both \politifact (blue) and \abc (orange). The distributions become roughly bell-shape and lightly skewed to the right for each dimension. %
Overall, the correlations values shown in figure for both individual (heatmaps in lower triangular matrix) and aggregated judgments (scatterplots in upper triangular matrix) are always positive, as it would be expected since all the seven dimensions share the same positive connotation.
Correlations are sometimes even quite high (e.g., $\rho=0.86$ between aggregated Correctness and Overall Truthfulness for \politifact statements), thus demonstrating some relations between different dimensions. However, some correlations are lower (e.g., $\tau=0.24$ and $0.2$ for Neutrality and Comprehensibility), thus highlighting a somehow higher independence between those dimensions.

\subsubsection{Internal Agreement Among Workers}

To measure the internal agreement, we used the Krippendorff's $\alpha$ \cite{krippendorff2011computing} metric both on the different ground truth level and at the unit level. %
The choice of using Krippendorff's $\alpha$ is  motivated not only by previous work \cite{la2020crowdsourcing,roitero2020crowd}, but also by theoretical reasons, since other agreement metrics are not suitable for our setting.
Cohen's $\kappa$ is used to compute agreement in the case of two assessors. Fleiss' $\kappa$, which generalizes Cohen's $\kappa$ to multiple assessors, can be only used when they assign categorical ratings, i.e., when they classify items. None of these can be applied to our case, where we have several assessors (i.e., 10) and an ordinal classification problem (i.e., the categories we consider are ranked). For these reasons, we used Krippendorff's $\alpha$ to compute the agreement with multiple assessors on non nominal scales. 
For a further analysis on agreement metrics see
\cite{fleiss1973equivalence,kvaalseth1989note,checco2017let}.

Results show that, overall, the agreement level is rather low. The $\alpha$ values for all the dimensions are in the $[.02,.08]$ range when computed for the statements all together, in the  $[-0.02,0.1]$ range when computed on the three \abc categories, and in the $[-0.02,0.1]$ range (with a mean value of $0.03$) for the \politifact categories, with the exception of the  \politifactbarelytrue and \politifacthalftrue categories which are in the $[-0.02,0.14]$ range (with a mean value of respectively $0.09$ and $0.05$).
It is known that $\alpha$ values are dependent on the amount of data and the evaluation scale considered \cite{checco2017let}. Given the fact that in our experiments both factors are fixed, this might be an indication that workers agree more when assessing statements on the middle of the truthfulness scale.

\subsubsection{External Agreement with Experts}

\begin{figure}[tb]
  \centering
  \begin{tabular}{@{}c@{}c@{}c@{}}
   \includegraphics[width=.33\linewidth]{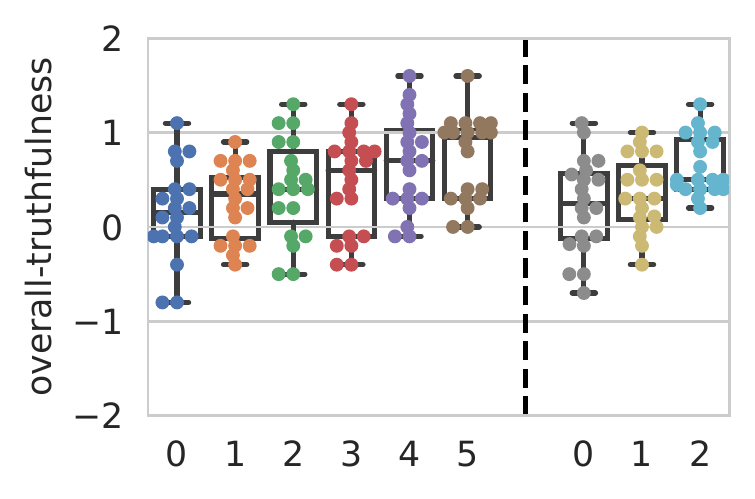}&
    \includegraphics[width=.33\linewidth]{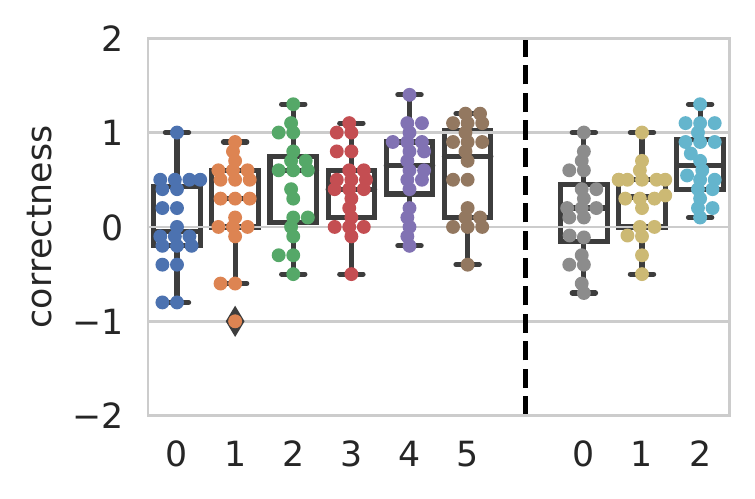}&
    \includegraphics[width=.33\linewidth]{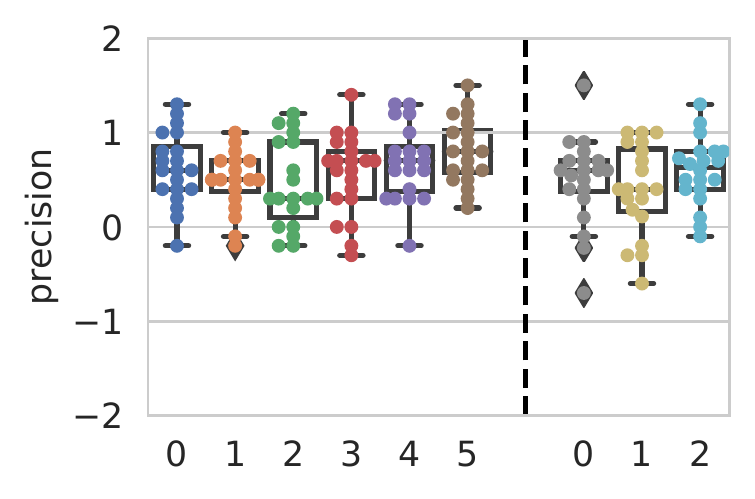}\\
  \end{tabular}
\caption{%
    Correlation with the ground truth of the Overall Truthfulness and behavior of the Correctness and Precision dimensions with a breakdown on \politifact and \abc labels. 
    Mean as aggregation function. When moving towards the right hand part of the plots, i.e., when going towards increasing truthfulness according to the ground truth, the median values are clearly increasing for Overall truthfulness (directly corresponding to the ground truth), but not necessarily so for the other dimensions (not directly related to the ground truth).
}
  \label{fig:scales-comparison}
\end{figure}

Moving from internal to external agreement, Figure~\ref{fig:scales-comparison} shows a plot of the workers scores aggregated with respect to the corresponding expert scores. Three dimensions are reported: Overall Truthfulness, Correctness, and Precision. 
Before commenting these plots, we make some remarks. 
First, it must be noted that the set of expert judgments is available only for the dimension named Overall Truthfulness, thus the remaining dimensions show the perceived value of the statements on each dimension with a breakdown on the \politifact and \abc categories. So, while Overall Truthfulness is meant to be correlated with experts' judgment, Precision captures an orthogonal and independent information. This is reflected by the different trend of the workers median scores reported in Figure~\ref{fig:scales-comparison}.
Second, the judgment scales used by the workers and by the experts on the Overall Truthfulness are slightly different; while experts provided their judgment on either a six (for \politifact) or a three level (for \abc) ordinal scale, the crowd workers provided their judgments on a five-level Likert scale. These scales are different both on the number of levels (six or three versus five), and also on the psychological interpretation of such scale. 

We now turn to analyze Figure~\ref{fig:scales-comparison}.
The ground truth is on the horizontal axis (\politifact on the left and \abc on the right) and the aggregated crowd judgments on the vertical axis. Each dot is a statement, the boxplots show the breakdown of the distributions (quantiles and median) for each ground truth level. 
Focusing on the plot on the left in Figure~\ref{fig:scales-comparison} we can see that on 
Overall Truthfulness increasing the ground truth level (i.e., going towards right in each plot) corresponds to an increasing judgment by the crowd. This is an indication that crowd workers provided judgments which are in agreement with the experts, despite the two set of judgments being on two different scales, both theoretically and psychologically.
We can also compare the correlation between Overall Truthfulness and the ground truth shown with the similar three plots shown in Figure~2 by \citet{roitero2020crowd} (one for each scale they used to collect truthfulness judgments). 
There is no noticeable qualitative difference, despite the judgments being again of different scales: overall, we can say that our crowd workers provided judgments of comparable quality to previous work.

Furthermore, the plots on the center and on the right of Figure~\ref{fig:scales-comparison} show that the specific dimensions of Correctness and Precision have a different appearance, and it can thus be considered somehow orthogonal to Overall Truthfulness.
\footnote{The other dimensions (not shown) show a  similar behavior to either Precision or Overall Truthfulness.}
We remark that we do not have any expert judgment for the dimensions with the exception of Overall Truthfulness, thus it does not make sense to directly correlate the other dimensions with the expert judgments, as each dimension can measure  different aspects from the ground truth (e.g., the Precision of a statement is not necessarily related to its Truthfulness). However, it might make sense to combine different dimensions to obtain a better measure of truthfulness, as we will discuss in the following sections.

To investigate the perceived disagreement between the expert and crowd judgments on Overall Truthfulness, and given that the two set of  judgments collected are on different scales, we computed how many times the aggregated values shown in the left plot of Figure~\ref{fig:scales-comparison} correspond to a value which is at the same distance between two values of the judgment scale used (i.e., the average is x.5, for x in the scale, $0 \leq x \leq 4$): this happens for about 20.5\% of statements. 
We compare this result  to each judgment scale used by \citet{roitero2020crowd}, since the set of statements is the same. When considering the three-, six-, and one hundred-level scales used by \citet{roitero2020crowd} the percentages of statements are very close, and  respectively of 19.4\%, 18.3\% and 23.9\%. This is an indication that the perceived disagreement between experts and crowd workers is not dependent on the scale used to collect the judgments, but it is attributable to other factors.

\begin{figure}[tbp]
  \centering
  \begin{tabular}{@{}c@{}c@{}c@{}c@{}c@{}}
   \includegraphics[width=.20\linewidth]{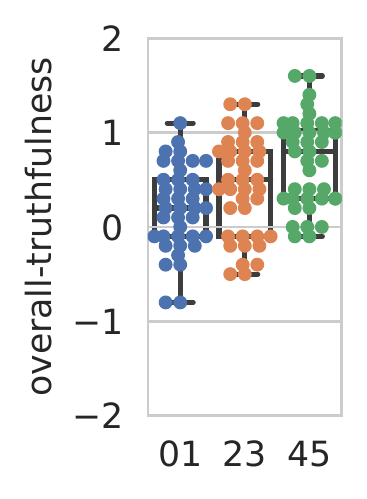}&
    \includegraphics[width=.20\linewidth]{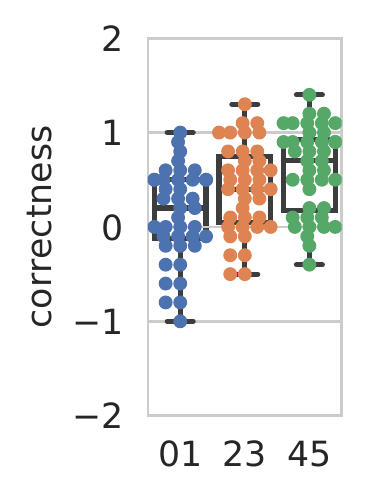}&
    \includegraphics[width=.20\linewidth]{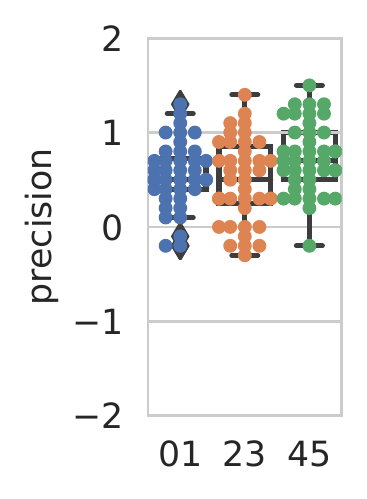}
    \includegraphics[width=.20\linewidth]{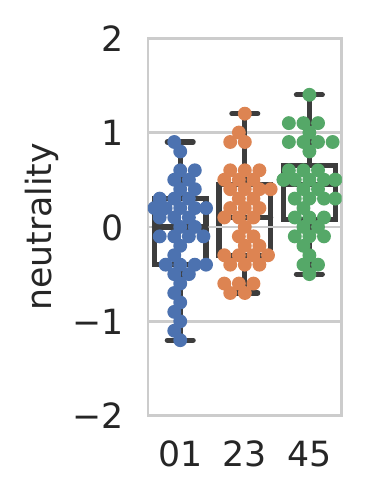}&
    \includegraphics[width=.20\linewidth]{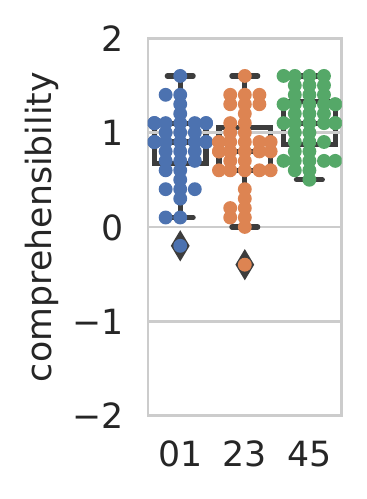}&
    \\
  \end{tabular}
\caption{%
    Correlation with the ground truth of Overall Truthfulness and a sample of the other dimensions.
    \politifact has been grouped into 3 bins.
    Mean as aggregation function. 
    The binning allows to see more clearly the increasing median trends; compare to Figure~\ref{fig:scales-comparison}.
}
  \label{fig:scales-comparison-bin3}
\end{figure}

In order to check if the agreement between experts and crowd workers can increase when considering a coarse grained scale, we grouped together ground truth levels, as done by \citet{roitero2020crowd}.
Figure~\ref{fig:scales-comparison-bin3} 
shows the correlation values between Overall Truthfulness and expert ground truth obtained by binning \politifact ground truth categories into 3 bins using mean as aggregation function. With respect to Figure~\ref{fig:scales-comparison} this binning allows to slightly improve some correlation values and to obtain a clearer trend of increasing median values for Overall Truthfulness. 
This result holds across each truthfulness dimension (the plots shows five of them) and is consistent with \citet{roitero2020crowd} findings.

\subsubsection{Behavioral Data}

\begin{figure}[tbp]
  \centering
  \begin{tabular}{@{}c@{}}
  \includegraphics[width=.5\linewidth]{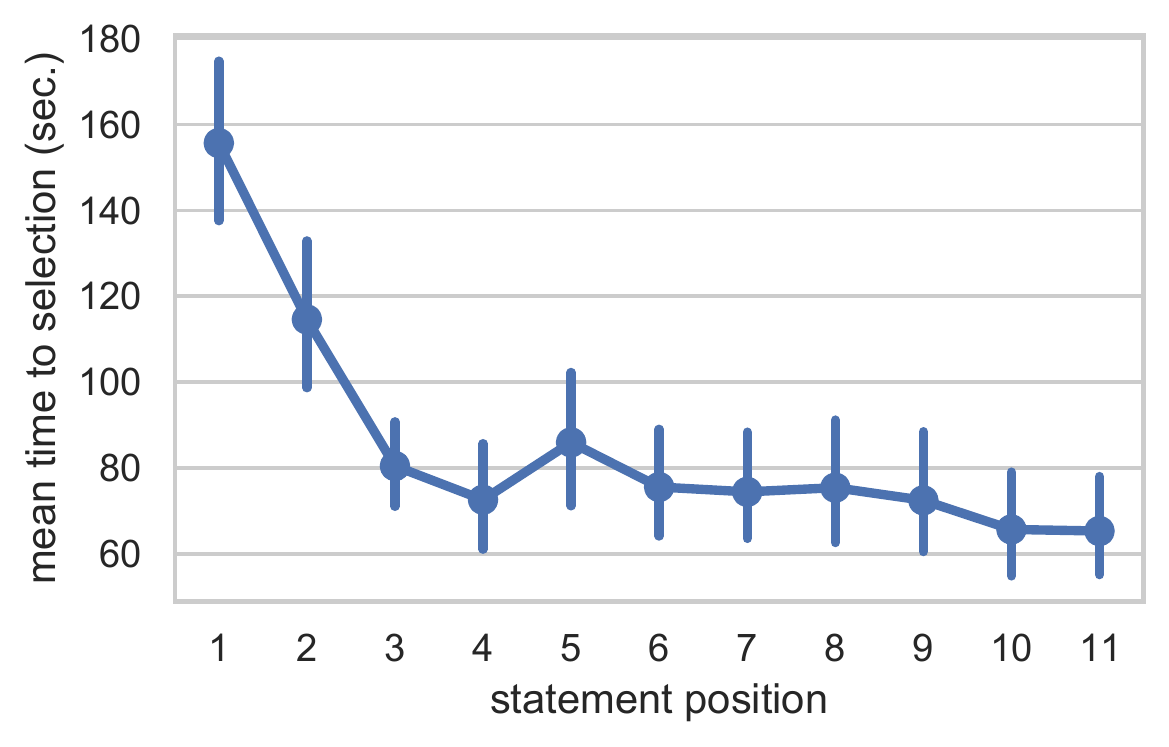}
  \end{tabular}
\caption{% 
Average time (in seconds) spent by workers to judge the Overall Truthfulness  for each statement position.
There is a clear learning effect as the worker goes through the task.
}
  \label{fig:dimensions-values-changes}
\end{figure}

We now turn to the analysis of workers's behavior while assessing each truthfulness dimension. Figure~\ref{fig:dimensions-values-changes} shows the average time spent by each worker to select a value for the Overall Truthfulness for each statement position. There is a clear indication of a learning effect since the average time spent to select a value for the Overall Truthfulness decreases while statement position increases. To support this finding, we also measured the statistical significance between the time values between each statement position. We found that the differences are statistically significant with a $p<.01$ level when considering positions $1$ and $2$ compared to any other position. When considering positions $3$ and $4$ there are statistically significant differences with a $p<.05$ level only with respect to the first two and the last two positions. These findings confirms that there is a learning effect within the first two positions which can last up to the fourth positions, and after the fourth statement the workers evaluate the subsequent statements in the same amount of time.

Workers spent most of the time assessing Overall Truthfulness because they were required to provide also a URL as justification for their choice. When considering other dimensions workers spent much less time to select a value and there are no clear trends visible. This is probably due to the fact that workers thought about the value to assign to other dimensions while assessing Overall Truthfulness. In more detail, the average time spent to select a value for other dimensions corresponds to 1.7 seconds for Confidence, 3 for Correctness, 4.1 for Neutrality, 5 for Comprehensibility, 6.2 for Precision, 7.1 for Completeness, 8.3 for Speaker’s Trustworthiness and 9.4 for Informativeness, much lower than the average time spent to assess the Overall Truthfulness, which is 85 seconds. 

\subsubsection{Summary}

Overall, from the analysis in this section we can draw several remarks about \ref{i:crowd-reliability}. Workers tend to agree with statements more than disagree, and since our dataset is balanced this holds also for false statements (Figure \ref{fig:correlation-between-dimensions}). 
Workers have on average a similar level of agreement on the set of statements they judge and an increasing ground truth level corresponds to increasing judgments by them, for Overall Truthfulness (Figures~\ref{fig:scales-comparison} and \ref{fig:scales-comparison-bin3}) and tend to agree more when assessing statements on the middle of the truthfulness scale. Workers learn how to assess the Overall Truthfulness (Figure \ref{fig:dimensions-values-changes}). 
These remarks let us conclude that workers put effort in providing quality judgments and these judgments are reliable and meaningful.

\subsection{\ref{i:rrq2}: Independence of the Dimensions}
\label{sect:truthfulness-dimensions}

The results reported so far show that the various dimensions, as well as Overall Truthfulness, are correlated to some extent. We now turn to understand if they anyway measure different aspects, or if some of them could indeed be derived from the other ones.
Going back to Figure~\ref{fig:correlation-between-dimensions}, one can find higher and lower correlations. 
The plots on the bottom left, concerning non-aggregated assessments, show higher correlations for Correctness with both Overall Truthfulness and Speaker’s Trustworthiness. 
The same is confirmed for aggregated assessments, shown on the top right, for which also Pearson's $\rho$ and Kendall's $\tau$ correlation values are included.
Focusing on the correlation of Overall Truthfulness with the seven dimensions (first row / first column) it appears clear that Neutrality, Comprehensibility, and Precision (0.48, 0.30, 0.43 $\tau$ respectively for aggregated judgments over \politifact statements and 0.31, 0.27, 0.30 $\tau$ for \abc statements) do not correlate well with Overall Truthfulness; Completeness, Speaker’s Trustworthiness, and Informativeness are slightly higher (0.53, 0.60, and 0.53 $\tau$ respectively for aggregated judgments over \politifact statements and 0.42, 0.56, 0.4 $\tau$ for \abc statements) but not as high as Correctness.
Summarizing, we can say that given a statement, each of the various dimensions measures a different aspect of truthfulness, and different from the Overall Truthfulness as well; this is true both when we look at individual worker assessments as well as at assessments aggregated over all workers who judged the same statement.

Reconsidering Figures~\ref{fig:scales-comparison} and~\ref{fig:scales-comparison-bin3} we find further confirmation of the independence of the dimensions, since it is true that all trends are similar, but there are also clear differences.
Seeking for further evidence, we did the following experiment. We employed the ANOVA analysis and we correlate the Overall Truthfulness as a function of the other dimensions. After fitting the ANOVA on such model, we used the $\omega^2$ index \cite{olejnikAnova} to measure the size of effect of each dimension in estimating the Overall Truthfulness. Results are as follows. 
The Overall Truthfulness score is mainly influenced (by one order of magnitude) by the 
Correctness ($\omega^2=0.228$), followed by
trustworthiness ($\omega^2=0.019$) and Informativeness ($\omega^2=0.017$). Comprehensibility ($\omega^2=0.008$), Completeness ($\omega^2=0.001$), Precision ($\omega^2=0$), and Correctness ($\omega^2=0$) have almost no effect.
We also fitted another ANOVA model to investigate the interactions between dimensions: results show that all interactions are weak ($\omega^2 \leq 0.04$) suggesting that indeed all the dimensions are somehow orthogonal and measure different aspects of the truthfulness of the statements. 
Nevertheless, the analysis of interaction between dimensions also shows that all dimensions are used by the workers when assessing the statements, and thus all dimensions are necessary (i.e., there is no redundant one). We leave for future work to investigate if other dimensions can be added to the existing ones in order to capture even more aspects when evaluating a statement. 

\begin{figure}[tbp]
  \centering
  \begin{tabular}{@{}c@{}c@{}}
  \includegraphics[width=.48\linewidth]{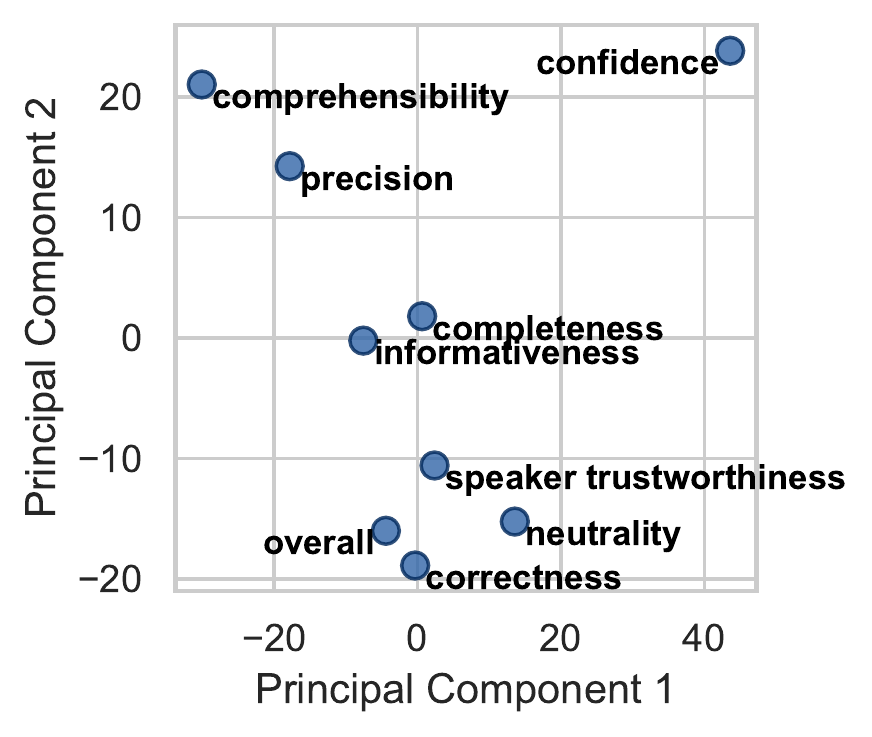}&
   \includegraphics[width=.48\linewidth]{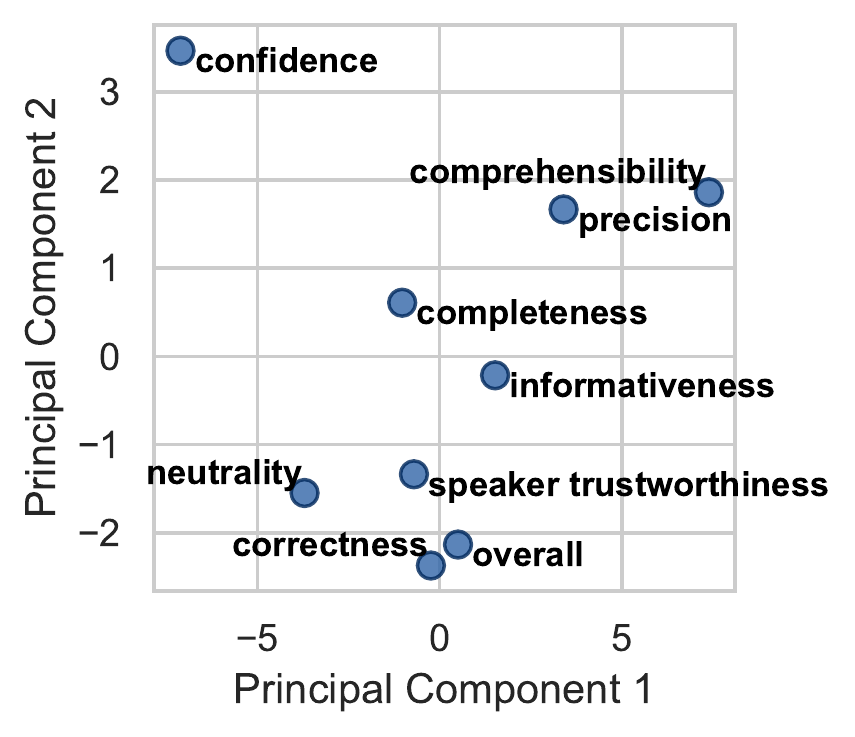}\\
  \end{tabular}
\caption{%
Principal components for the statements $\times$ judgments matrix: individual (left) and aggregated (right) judgments.
The plots show the relative similarity of one dimension to the others; the most similar dimensions to Overall Truthfulness are Correctness, Speaker Trustworthiness, and to a lesser extent, Neutrality.
}
  \label{fig:clustering}
\end{figure}

To further study the relationships and independence of dimensions we performed the following experiment. We considered both the individual and aggregated judgments to build a statement $\times$ judgments matrix. Then, we computed the Principal Components Analysis (PCA) of such matrix, with the aim of finding the orthogonal bases which explain the maximal variance of data. In the computed space with the new coordinate system, we considered the two components (i.e., dimensions) which explain the majority of the variance. 
Figure~\ref{fig:clustering} shows the result of the PCA analysis on the individual (left) and aggregated (right) judgments. As we can see from the plots, especially focusing on the position of the other dimensions with respect to Overall Truthfulness, is that the most similar dimensions to Overall Truthfulness are Correctness, Speaker trustworthiness, and to a lesser extent, Neutrality. This behavior holds for both the individual and aggregated judgments. It makes sense that when a worker provides a judgments for the Overall Truthfulness of a statement, the dimensions which are more correlated with its judgments are the ones identified by the PCA analysis. 
On the contrary, from Figure~\ref{fig:clustering} we see that other dimensions, such as Confidence, Comprehensibility, and Precision, are not related to any other dimension and are the most distant from Overall Truthfulness as well; again, this behavior perfectly makes sense  thinking about the process of assessing the truthfulness of a statement. In future work we plan to conduct a study to investigate if the same behavior is present in the experts judges.
Summarizing, the PCA analysis confirmed that all dimensions are needed and different, and allowed us to draw meaningful information on the relationships and similarities between those dimensions.

We now turn to study whether it is possible to combine the individual dimensions in a way that it improves agreement between the crowd and expert judgments.

\begin{figure}[tbp]
  \centering
  \begin{tabular}{@{}c@{}c@{}}
  \includegraphics[width=.48\linewidth]{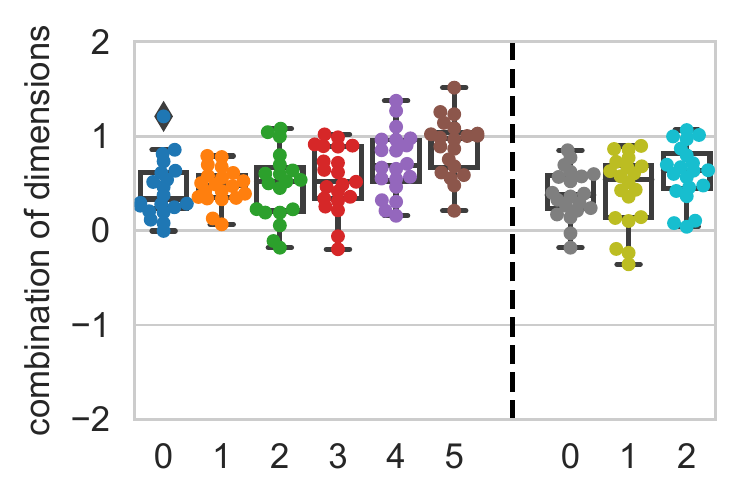}&
   \includegraphics[width=.48\linewidth]{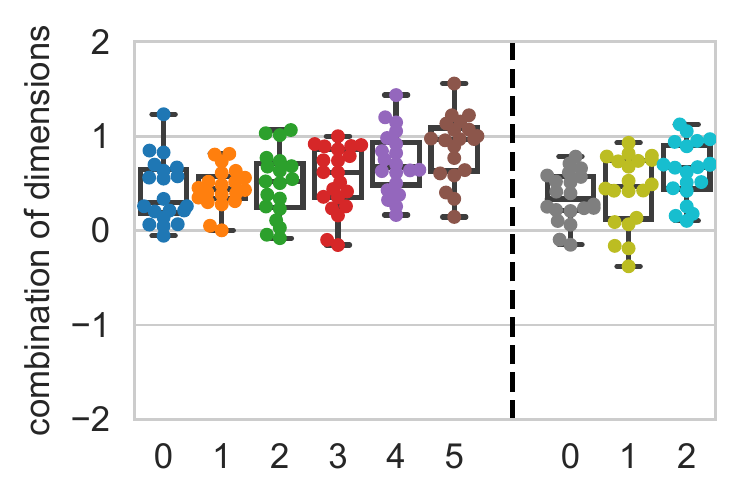}\\
  \end{tabular}
\caption{%
Truthfulness dimensions first aggregated using the mean function, then
combined, using either the $\omega^2$ values (left), or the CRT scores (right).
Combining the dimensions still allows to obtain increasing median values when moving towards higher truthfulness, but it does not seem to improve results from Figure~\ref{fig:scales-comparison}.
}
  \label{fig:scales-comparison-combination}
\end{figure}

Since  the individual dimensions measure different aspects, we could hypothesize that a combination of the assessments on certain individual dimensions could lead to a better approximation of the ground truth than using the Overall Truthfulness only.

The judgments collected for each truthfulness dimension can be combined together and used to predict the ground truth categories for both \politifact and \abc. To do so, we employ the ANOVA analysis using the $\omega^2$ index  to estimate the size of effect of each dimension when used to estimate the ground truth; note that the ground truth values for the statements is not available in the real setting, thus we are estimating the combination of dimensions in a sort of ideal scenario.
After we computed the $\omega^2$ index of each dimension, we aggregate the 10 judgments for each statement using the weighted mean function, where the weights are the $\omega^2$ values.
Figure~\ref{fig:scales-comparison-combination} left plot shows the correlation values between the label obtained by combining each dimension and ground truth categories. 
Overall, we can say that combining the dimension still allows to obtain increasing median values when moving towards higher truthfulness values,
but it does not seem an improvement of the left plot in Figure~\ref{fig:scales-comparison}. 

As another approach we tried to exploit the CRT answers.
First, all the judgments are aggregated using weighted mean where the weights are the ratio of correct answers given by each worker to the CRT questions normalized in $[0.5,1]$ interval (i.e., we weight more the judgments from high quality workers). 
Then, all the dimensions are combined using a weighted mean function where the weights are the $\omega^2$ scores computed above.
Results are shown in Figure~\ref{fig:scales-comparison-combination} right plot. There is no significant difference with respect to the aggregation shown in Figure~\ref{fig:scales-comparison-combination} left plot. 
 
To better understand this somehow negative results in the combination of dimensions, we employed again the ANOVA analysis.
In more detail, we fitted two ANOVA models: in the former we correlate the ground truth values to the all the dimensions, in the latter we correlate the ground truth values to the Overall Truthfulness dimension alone. Results show that 
the residual in both cases is very similar, indicating that there is no major difference when trying to predict the ground truth label using the Overall Truthfulness alone or a combination of all the dimensions.\footnote{
Similar analyses have been proposed to understand the contributions of each component to the quality of a system \cite{ferro2016general,ferro2018toward,ferro2019using,zampieri2019topic,roitero2020leveraging}.} The $\omega^2$ index for the latter model is rather low (i.e., $0.02$), indicating that indeed the Overall Truthfulness dimension alone is not sufficient to predict the ground truth label, neither are naive combinations of the dimensions.
It seems that an effective combination of dimensions cannot be achieved by simple models. 
We leave for future work more complex approaches such as hierarchical models (that might require a modification in the experimental analysis), the combination of dimensions by means of complex (e.g., non linear) functions, or even the exploitation of other data as the URL provided. We will also consider requesting additional information required to the worker, such as a confidence value and a textual justification for each dimension, which will probably require a slightly different experimental design to avoid to overload the worker.

\subsection{\ref{i:worker-behavior}: Worker Behavior}\label{sect:worker-behavior}

Considering the still inconclusive results from the combination of dimensions, we  also did a first attempt to consider the worker behavior, as a proxy for worker quality,
to boost the correlation values between the collected judgments and the ground truth. The simple idea is to give more weight to the workers with higher quality, and to use the CRT answers to estimate worker quality.

\begin{figure}[tbp]
  \centering
  \begin{tabular}{@{}c@{}c@{}}
   \includegraphics[width=.48\linewidth]{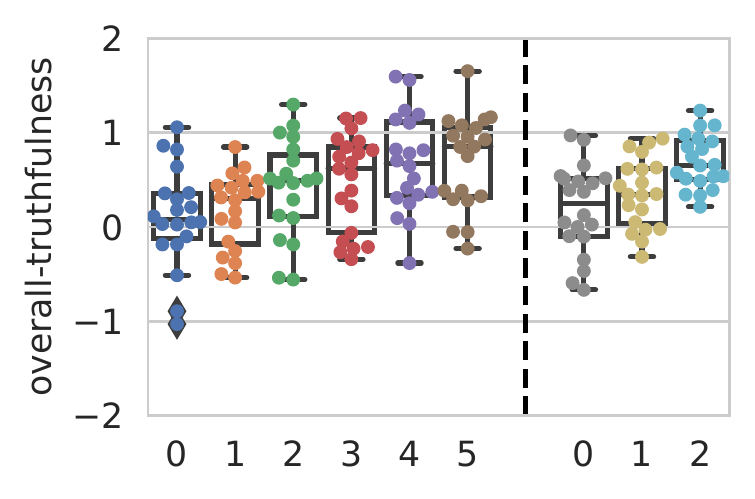}&
    \includegraphics[width=.48\linewidth]{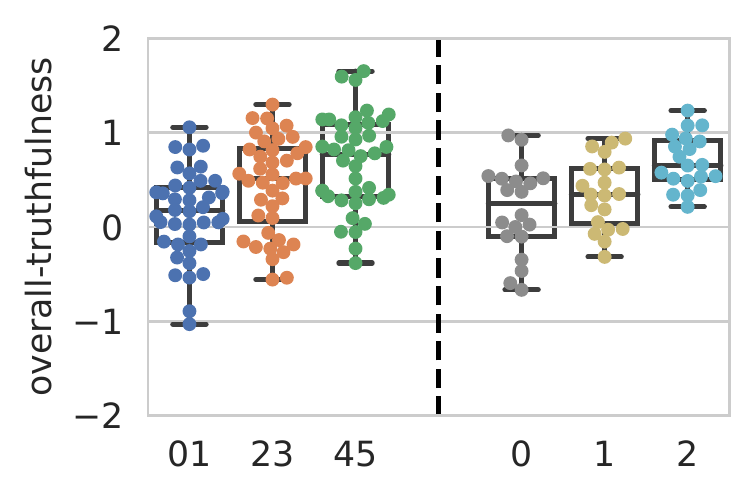}
  \end{tabular}
\caption{%
Overall Truthfulness judgments aggregated with the weighted mean function using the CRT scores (left); \politifact categories are grouped together (right).
This approach does not improve the correlation with the ground truth: compare with Figures~\ref{fig:scales-comparison} and~\ref{fig:scales-comparison-bin3}.
}
  \label{fig:agg-crt}
\end{figure}

Workers which answered correctly to all three CRT questions are 18\%; another 18\% answered correctly to 2 questions, 18\% to 1 question, and 34\% did not answer correctly to any question. 
We aggregated the individual judgments with the weighted mean function, using as weights the normalized CRT scores: for each worker, we considered the amount of correct answers (out of 3) for the CRT questionnaire and we normalized the score in the $[0.5,1]$ range.
Figure~\ref{fig:agg-crt} left plot shows the correlation of  the Overall Truthfulness values obtained by such a weighted mean with \politifact and \abc ground truth; right plot shows the result when grouping the categories into 3 bins. As we can see, the resulting plots are very similar to the top left plots in Figures~\ref{fig:scales-comparison} and~\ref{fig:scales-comparison-bin3}, thus it seems that this approach does not improve the correlation with the ground truth.
We plan to investigate more complex worker behaviors and their relations with aggregation functions in future work.

We also remark that when considering the individual (i.e., not aggregated) judgments for each statement without gold questions, the majority of workers tend to use distinct labels to provide his/her judgment. Without considering self-reported confidence, each worker provides 8 judgments by choosing labels from a set of five possible values. Only 12\% of workers used the same label for all dimensions, whereas 29\% used two distinct labels, 39\% used 3 distinct labels, 18\% used 4 distinct labels, and 2\% used all 5 distinct values. The majority of workers tends to use most of the judgment scale to provide their judgments.   
This is another confirmation of dimensions independence (\ref{i:rrq2} and Section~\ref{sect:truthfulness-dimensions}) and shows how different dimensions cover different aspects of truthfulness.

\subsection{\ref{i:dim-informativeness}: Dimension Informativeness}
\label{sect:informativeness}
We now evaluate the informativeness of the multidimensional assessments (not to be confused with the truthfulness dimension  called Informativeness  used to evaluate the statements).
First, we test whether it is possible to synthesize these judgments computationally, and the two dimensions for which we found computational counterparts are Comprehensibility and Correctness. 
Readability measures determine the understandability of text which might affect Comprehensibility. 
We compute the readability of all the statements for 10 measures: 
Flesh Kincaid Reading Ease, Flesh-Kincaid Grade Level, Automated Readability Index, Gunning Fog Index (see ~\citet{kincaid1975derivation});  Dale-Chall (see ~\citet{dale1948formula}); Simple Measure of Gobbledygook (SMOG, see~\citet{mclaughlin1969smog}); Coleman-Liau Index (see~\citet{coleman1975readability}); 
Forcast (see~\citet{caylor1973methodologies}); and Lesbarhets Index, Rate Index (LIX, RIX, see~\citet{biornsson1968lasbarhet}). All of them show a low correlation with the Comprehensibility scores (with a maximum $\rho=0.19$ for RIX). We conclude that the information provided by the workers with the Comprehensibility scores is hardly captured by automated readability scores, and thus it is a significant measure to be crowdsourced.
We also compare the Correctness scores with the statement polarity computed using the Textblob Python library.\footnote{\url{https://textblob.readthedocs.io/en/dev/index.html}.} However, polarity measures the statement emphasis, while Correctness focuses on the content level. As a result, their correlation is weak ($\rho=0.13$).

Second, we investigate how each single dimension judgments can contribute to understand the motivations behind the overall judgment \cite{lawrence2020argument} as follows.
For \abc statements, our ground truth provides also an assessment rationale (e.g., ``cherry picking''). We compute the Word Mover's Distance ($\mathrm{wmd}$) \cite{wmd} between each rationale and the name of each dimension, and we check whether it correlates with the scores of that dimension. Consider the case where we have two statements, \textbf{statement$_i$} and \textbf{statement$_j$}, their \textbf{Precision} scores are 2 and 1 respectively, and their ground truth rationales are ``exaggeration'' and ``wrong''. In such a situation, we compute the correlation between the two scores (i.e., 2 and 1) and the semantic similarity of the word pairs (rationale, dimension name):
\begin{equation*}
    \mathrm{corr}((2,1),(\mathrm{wmd}(\mathit{exaggeration}, \mathit{precision}),\mathrm{wmd}(\mathit{wrong},\mathit{precision}))).
\end{equation*}
The scores show a weak correlation with the semantic distance between the labels and the corresponding dimension name (with a peak at 0.3 Pearson's $\rho$ correlation for Informativeness). However, combinations of similarity scores and metrics scores show a higher correlation (e.g., Overall Truthfulness values vs. Informativeness similarity 0.38, Speaker’s Trustworthiness vs. Completeness similarity 0.3). These preliminary insights indicate that the dimensions scores can help identifying the motivation behind the overall assessment of a statement. The combinations of similarities and scores will be further investigated in the future.

\subsection{\ref{i:machine-learning}: Learning Truthfulness from Multidimensional Judgments}

In this section we use a machine learning based approach to analyze the usefulness of the multidimensional assessments and of the worker behavior in supporting the prediction of expert judgments, both for \politifact and \abc.
We take two approaches here. First, we evaluate a number of supervised approaches in being able to predict the exact truthfulness verdicts provided by experts. Second, we use unsupervised and hybrid approaches to estimate truthfulness scores that are semantically close to the ground truth.

\subsubsection{Supervised Approach}
\label{sec:sup_app}
We aim to predict
the \politifact and \abc judgments, considering for \abc both the three-level scale and the original verdicts, with 30 different labels in our sample. The latter is the scale  initially used by experts when assessing the truthfulness of a statement and it is semantically more informative than the simplified one.
We considered the following features, computed for each judgment.
The one-hot-encoding of the worker ids in order to identify which worker performed the judgments, 
followed by the worker judgments on all the dimensions, and the 300-dimensional embedding\footnote{We consider the \texttt{SISTER (SImple SenTence EmbeddeR)} implementation, see \url{https://github.com/tofunlp/sister}.} of the string obtained from the concatenation of the query issued by the worker, and the title, snippet, and domain of the URL selected by the worker. The rationale behind this set of embeddings is that we try to capture the semantic relationship between the expert classification and the piece of information used by the worker to justify its judgment.

After computing the features, we divided our dataset into training and test sets. To avoid any possible bias or overfitting we compute the effectiveness metrics over 3 folds obtained using stratified sampling. 
We considered the following baselines. The first (i.e., ``Most Frequent'') predicts always the most frequent class present in the training set; the second (i.e., ``Weighted Sampling'') predicts, for each instance in the test set, a weighted random choice among the classes present in the training set, where the weights are the frequencies of each class; we repeat the process for the second baseline \num{1000} times for each fold. Finally, the third baseline (i.e., ``Random Choice'') simply returns a random class.
Apart from the three baselines, we employ the following supervised classification algorithms:
Random Forest,
Logistic Regression,
AdaBoost,
Naive Bayes, and 
Support Vector Machine (SVM).\footnote{We use the \texttt{sklearn} implementation of the algorithms, see \url{https://scikit-learn.org/stable/supervised_learning.html\#supervised-learning}.}
The parameters used to train the algorithms, reported to allow reproducibility, can be found in the repository containing the dataset that we release.

\begin{table}[tb]
\caption{
Effectiveness metrics when predicting the expert judgment. Baselines above the dashed line.
Random Forest is significantly more effective than the best baseline. 
    \label{tab:effectivneness-abc-prediction}}
\centering
\adjustbox{max width=0.75\textwidth}{%
\begin{tabular}{l cccc}
\toprule
\textbf{Algorithm} & \textbf{Accuracy} & \textbf{Precision} & \textbf{Recall} & \textbf{F1}  \\
\midrule
\multicolumn{5}{c}{\textbf{\politifact 6 Levels}}\\
\midrule
Random Choice                   & .167 & .167 & .167 & .167 \\
\hdashline
Random Forest                   & \textbf{.556} & \textbf{.561} & \textbf{.556} & \textbf{.554} \\
Random Forest (bootstrap CI)     & $[.477,.569]$&$[.482,.574]$&$[.477,.569]$&$[.476,.568]$ \\
Logistic Regression             & .391 & .417 & .392 & .392 \\
AdaBoost                        & .327 & .340 & .327 & .327 \\
Naive Bayes                     & .165 & .185 & .165 & .064 \\
SVM                             & .225 & .213 & .226 & .207 \\
  \midrule
\multicolumn{5}{c}{\textbf{\abc 3 Levels (Simplified)}}\\
\midrule
Random Choice                   & .333 & .333 & .333 & .333 \\
\hdashline
Random Forest                   & \textbf{.667} & \textbf{.670} & \textbf{.667} & \textbf{.665} \\
Random Forest (bootstrap CI)     &$[.594,.716]$&$[.595,.720]$&$[.594,.716]$&$[.592,.715]$ \\
Logistic Regression             & .557 & .563 & .557 & .555 \\
AdaBoost                        & .560 & .562 & .560 & .559 \\
Naive Bayes                     & .579 & .584 & .579 & .576 \\
SVM                             & .392 & .391 & .392 & .379 \\
\midrule
\multicolumn{5}{c}{\textbf{\abc 30 Levels (Original)}}\\
\midrule
Random Choice                   & .033 & .033 & .033 & .033 \\
Most Frequent                   & .134 & .018 & .134 & .032 \\
Weighted Sampling               & .067 & .067 & .067 & .066 \\
\hdashline
Random Forest                   & \textbf{.518} & \textbf{.562} & \textbf{.518} & \textbf{.491} \\
Random Forest (bootstrap CI)     & $[.426,.538]$&$[.460,.605]$&$[.426,.538]$&$[.398,.514]$ \\
Logistic Regression             & .195 & .151 & .195 & .143 \\
AdaBoost                        & .148 & .088 & .148 & .073 \\
Naive Bayes                     & .203 & .221 & .203 & .181 \\
SVM                             & .154 & .052 & .154 & .075 \\
\bottomrule
\end{tabular}
}
\end{table}

Table~\ref{tab:effectivneness-abc-prediction} reports the effectiveness scores obtained when predicting the  \politifact and \abc verdicts.
To deal with class imbalance, we report the weighted-averaged version of the Precision, Recall, and F1 metrics, i.e., we aggregate the effectiveness scores of all classes weighted by their frequency.
As we can see from the table, the Random Forest algorithm is able to predict the expert verdict better than both the random baselines and the other algorithms, for all the datasets considered.
To investigate the reason behind the differences in effectiveness between Random Forest and the other algorithms, we investigated the importance of the features used by the algorithm,\footnote{See \url{https://scikit-learn.org/stable/auto_examples/ensemble/plot_forest_importances.html}}
and we found that Random Forest considers equally all the features in the embedding vector, which are the most important for such algorithm; the rest of the features (i.e., the one hot encoding of the worker ids and the worker judgments) have an importance which is lower than the embedding vector, but still present; as evidence of that, if we remove either the workers id vector or the judgments, the effectiveness metrics decrease.
Thus, it seems that Random Forest is able to use all the input features to correctly classify the training instances, and to effectively generalize to novel ones.
This is an important result, as it indicates that multiple signals from the workers, namely their search sessions, can be leveraged to successfully predict the expert verdicts. It is important to notice that this is also true for the 30-class scenario of the original---and more semantically meaningful---\abc verdicts.

\begin{figure}[tbp]
  \centering
  \begin{tabular}{@{}c@{}}
   \includegraphics[width=.99\linewidth]{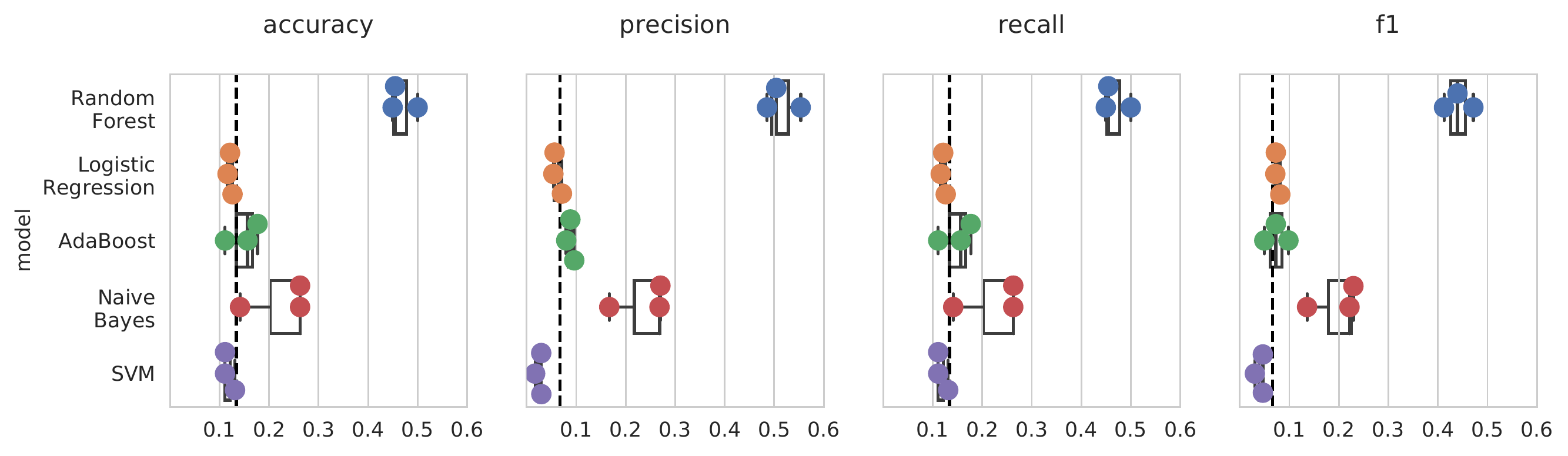}
  \end{tabular}
\caption{%
Effectiveness over the 3 folds for the \abc 30 levels case. The dashed line represents the best baseline. The effectiveness of the Random Forest classifier is clearly higher than the best baseline, even if statistical significance does not hold.
}
  \label{fig:stat-sign}
\end{figure}

We also investigated the statistical significance of the metric scores when comparing them against the best baseline; to this aim, we used the Wilcoxon signed-rank test\footnote{See \url{https://docs.scipy.org/doc/scipy/reference/stats.html}.} (paired data, non parametric test) and we corrected the results for multiple comparisons using the Bonferroni correction; we found that none of the comparisons are statistically significant, and all have $p>0.05$; this is most likely due to the low number of data points considered in the test (i.e., 3 since we split the data using 3 folds). As a further analysis, we plotted for each effectiveness metric the scores for the different folds, and we highlighted the best baseline with a dashed line (note that the baseline always obtains the same effectiveness score for all the folds). 
Figure~\ref{fig:stat-sign} shows the results. As we can see from the plots, it is reasonable to assume that the best performing algorithms are significantly better than the best baseline even though the statistical significance does not hold.
The same behavior holds for the \politifact 6 levels and \abc 3 levels case (not shown). As a final analysis, we employed the bootstrap technique to compute the 95\% confidence interval for the most effective algorithm (i.e., Random Forest); we employed \num{100000} stratified samples, and we computed the 2.5-th and 97.5-th percentiles \cite{linnet2000nonparametric,bland2015statistics}, in order to compute the 95\% likelihood that the computed range covers the true statistic mean. The results are shown in  Table~\ref{tab:effectivneness-abc-prediction}; as we can see, even considering the 2.5-th percentile, Random Forest is significantly more effective than the best baseline.

Given that the purpose of this paper is to study the impact of using a multidimensional scale, we investigated the performances of the machine learning techniques when different sets of dimensions are used with the aim of predicting the \politifact and \abc judgments. In more detail, we trained the same algorithms considered in Table~\ref{tab:effectivneness-abc-prediction} by using three groups of features:
(i) all the dimensions apart from Overall Truthfulness, 
(ii) only the Overall Truthfulness dimension, and 
(iii) all the dimensions and Overall Truthfulness.
Results (not shown) are almost indistinguishable from the ones of Table~\ref{tab:effectivneness-abc-prediction}, with very small fluctuations. Nevertheless, we found that it is almost always the case that the effectiveness metrics obtained when training the algorithms with (i) all the dimensions apart from Overall Truthfulness are little higher than the ones obtained when training considering (iii) all the dimensions and Overall Truthfulness; both approaches lead to obtain higher effectiveness metrics than the ones obtained considering (ii) only the Overall Truthfulness dimension.
As before, we also investigated the statistical significance between all the pairs of approaches using the Wilcoxon signed-rank test and correcting for multiple comparisons; we found that all differences are not statistically significant.
Summarizing, our results indicate that using all the dimensions to train a supervised approach leads to obtain the best (even though not significant) effectiveness metrics; we also found that Overall Truthfulness does not provide a significant improvement when used as a feature, and is outperformed when all the other dimensions are used.%

\subsubsection{Unsupervised Approach}
In addition to using a supervised approach as above described in Section \ref{sec:sup_app},  
we evaluate here the use of unsupervised approaches for truthfulness prediction.
Considering both supervised and unsupervised approaches gives us a complete overview of the expected effectiveness of the methods that can be used to predict a given verdict.% 
Our goal  is to predict a verdict that is semantically close to, and which polarity agrees with, the ground truth. However, we do not aim at predicting the exact label used in the ground truth. In particular, we focus on the \abc verdicts, which are semantically rich. %
This analysis helps us in understanding the links and relationships between the expert judgments and the workers assessments. In particular, we aim to understand if the weighted embeddings derived from the workers judgments only are aligned with the judgments produced by the experts. %  
Thus, we evaluate our predictions by checking Word Mover's semantic distance and sentiment difference.%
\footnote{We compute sentiment scores using Flair: \url{https://github.com/flairNLP/flair}}
Sentiment scores range between -1 and +1, and while semantic similarity tells us whether the rationale for the judgments are similar, sentiment difference tell us whether the polarities agree (e.g., comprehensible and accurate have a higher semantic distance than comprehensible and incomprehensible, but the sentiment difference is higher in the second case).
We compare our results with the worst, best, and average combinations obtainable by picking judgments in our ground truth.
If we picked a random verdict from the ground truth for each statement, we would obtain 
an average semantic distance of 2.48 in the best scenario, and of 4.41 in the worst. The average distance from random judgments is 3.40. Also, the worst possible sentiment difference is 1.97 and the best (excluding the case when we pick the exact right judgment) is 0.02. The average sentiment difference is 1.00. 
Here we focus on the statements, considering the average value of the assessments given by the workers.
We describe our strategies as follows.

\textit{Weighted Average Word Embeddings.}
We start from the assumption that our quality dimensions are positively connoted: when a worker assigns a +2 score to comprehensibility, we assume the overall verdict to imply that the statement is comprehensible. So, we compute the word embedding of each dimension name, and we weight it on the basis of the corresponding score. Then, we average the resulting embeddings to obtain an expected representation of the verdict's embedding. We lookup in the embedding dictionary the term 
having the closest embedding to this average embedding. 
The resulting labels have an average semantic distance from the ground truth of 4.14 and an average sentiment difference of 1.31: our
performance does not improve the random selection of judgments from the ground truth. This is also because, while the ground truth judgments belong to the same semantic area of quality assessment, our method searches the whole embedding dictionary.

\begin{figure}[tbp]
   \centering
   \begin{tabular}{@{}c@{}c@{}c@{}}
   \includegraphics[width=.33\linewidth]{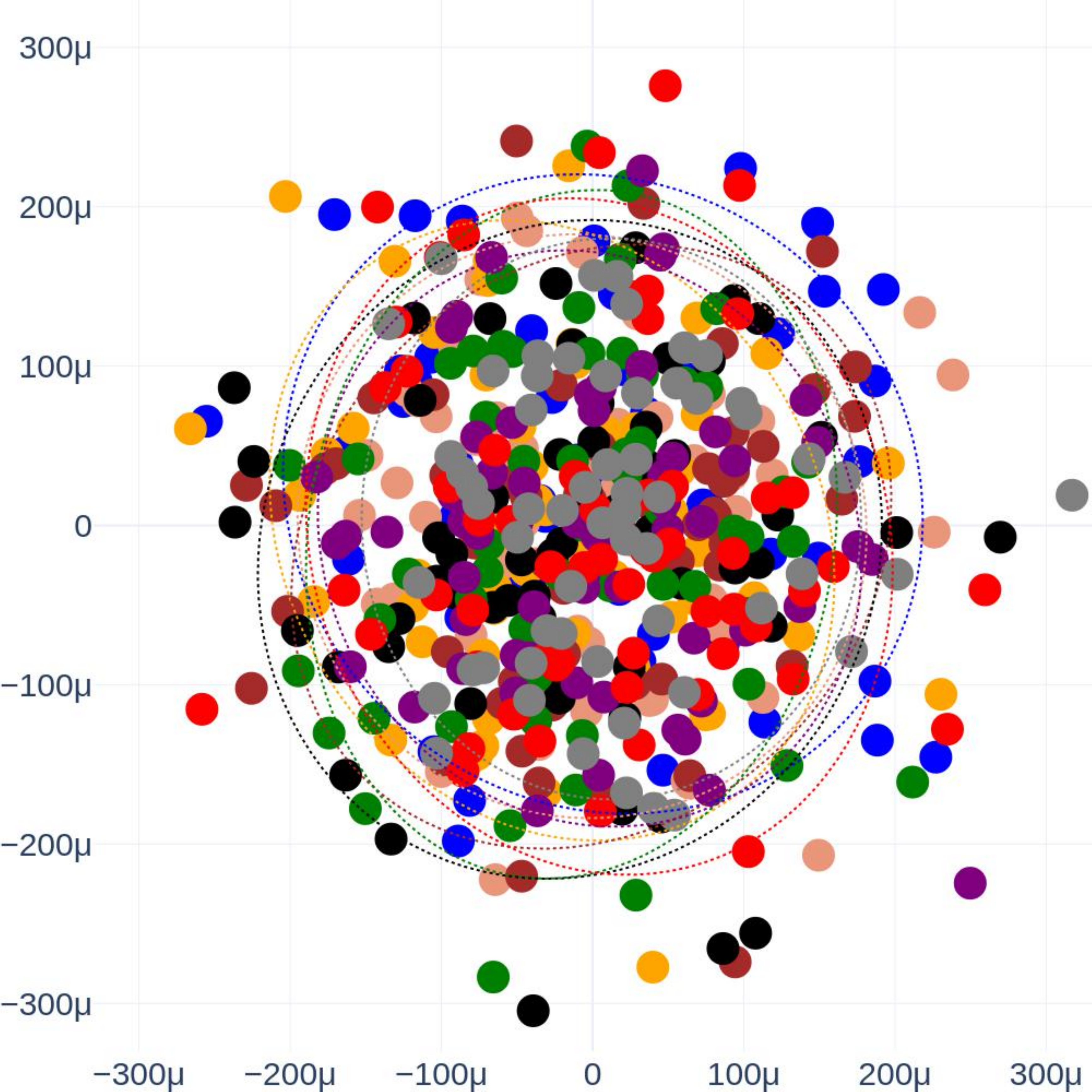}&
   \includegraphics[width=.33\linewidth]{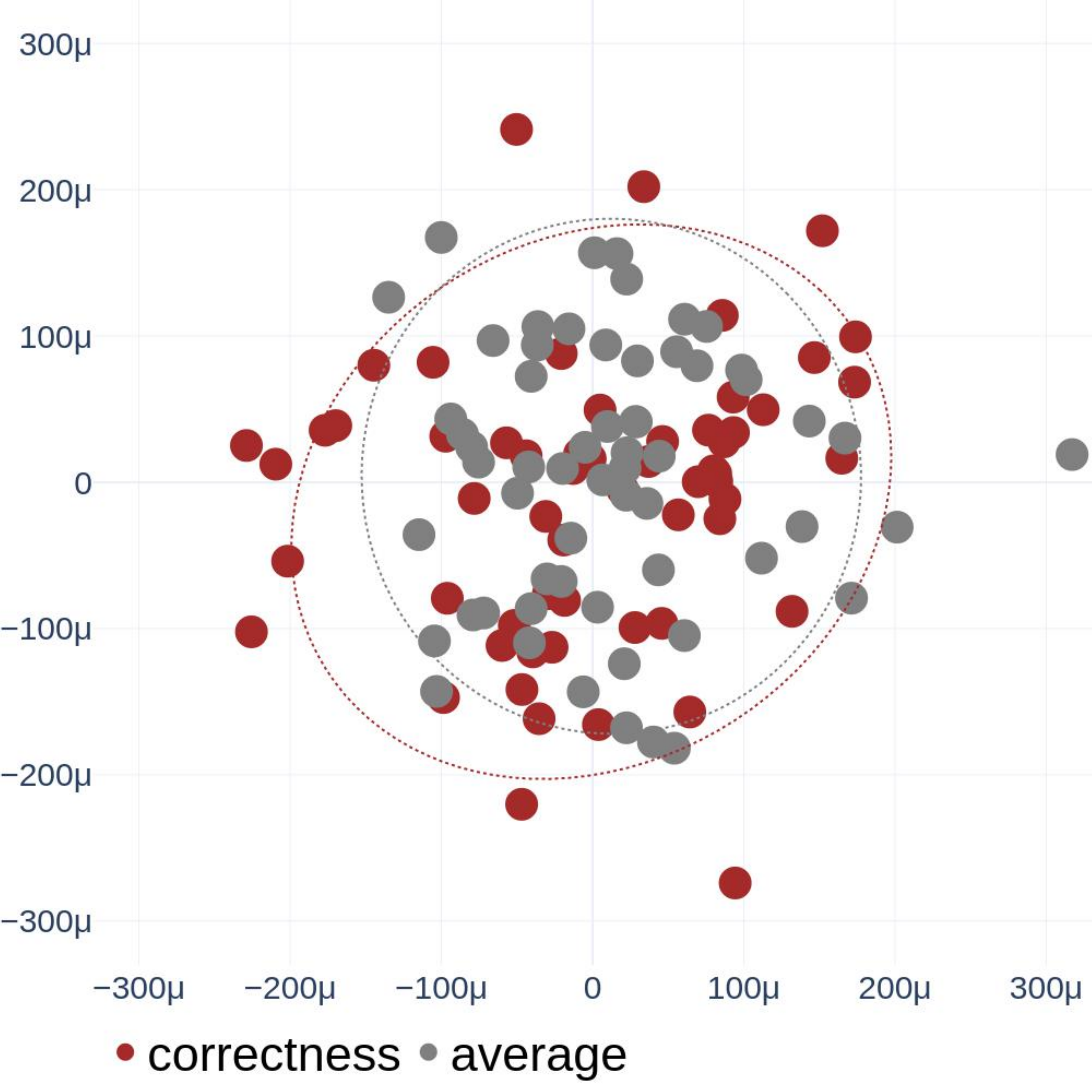}&
   \includegraphics[width=.33\linewidth]{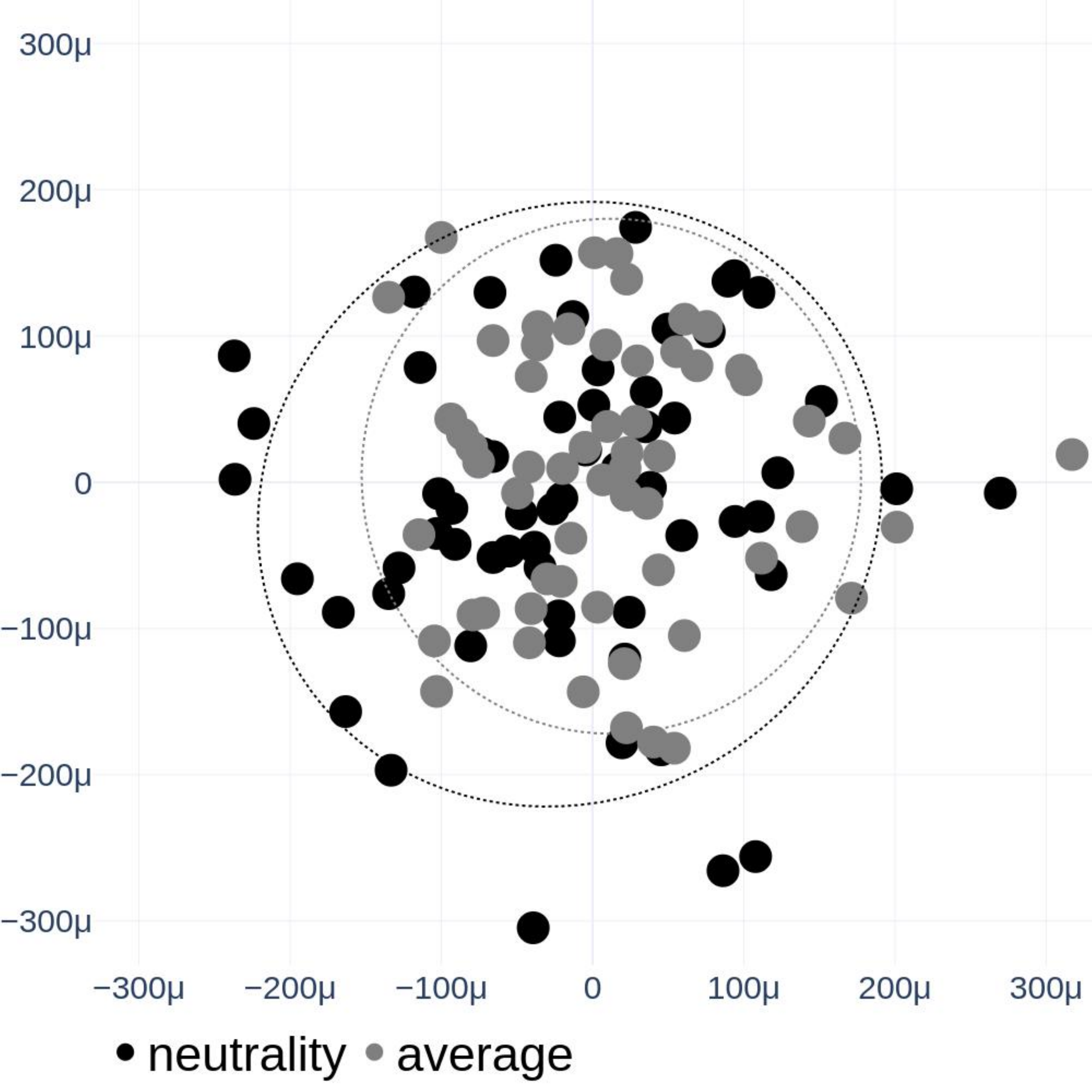}\\
   \includegraphics[width=.33\linewidth]{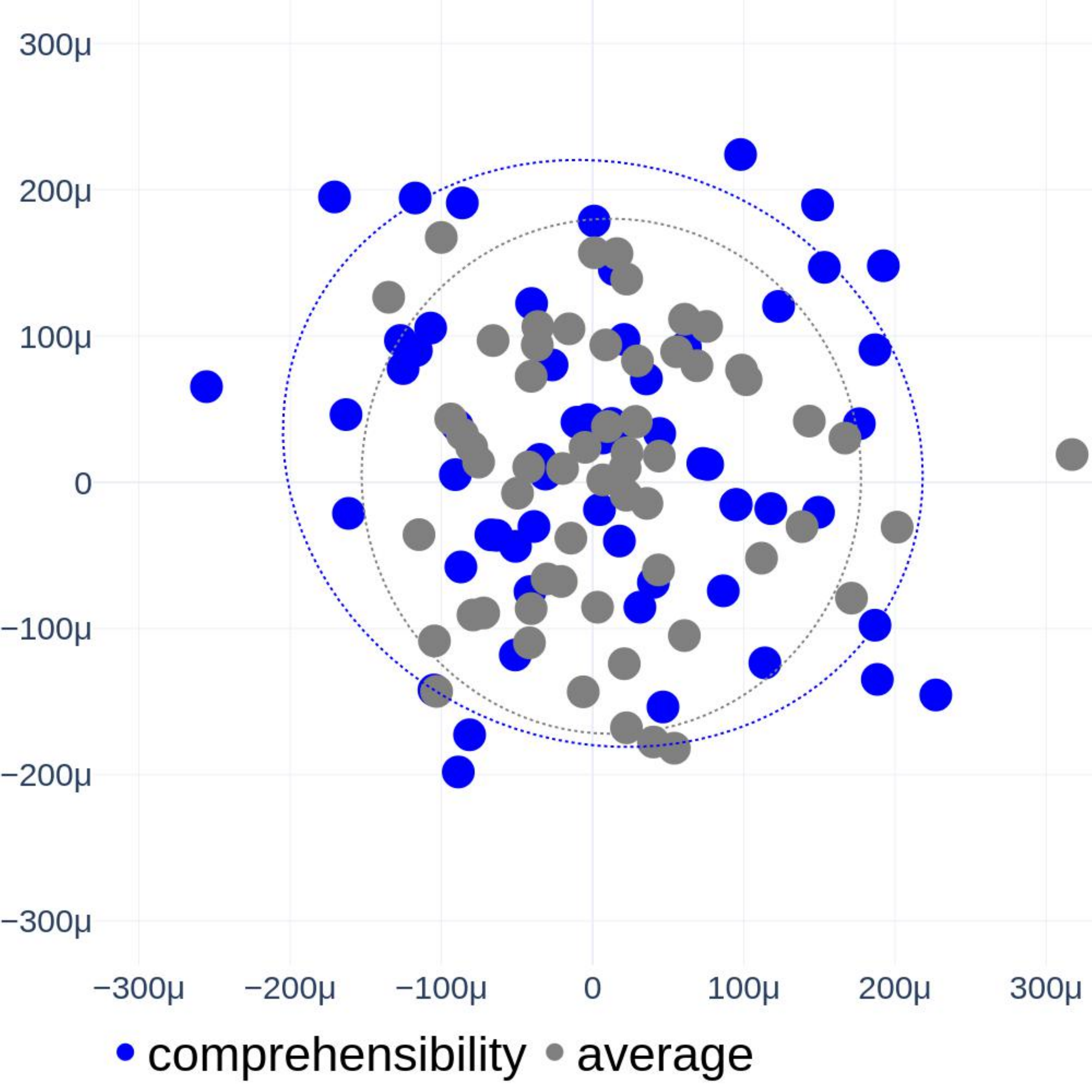}&
   \includegraphics[width=.33\linewidth]{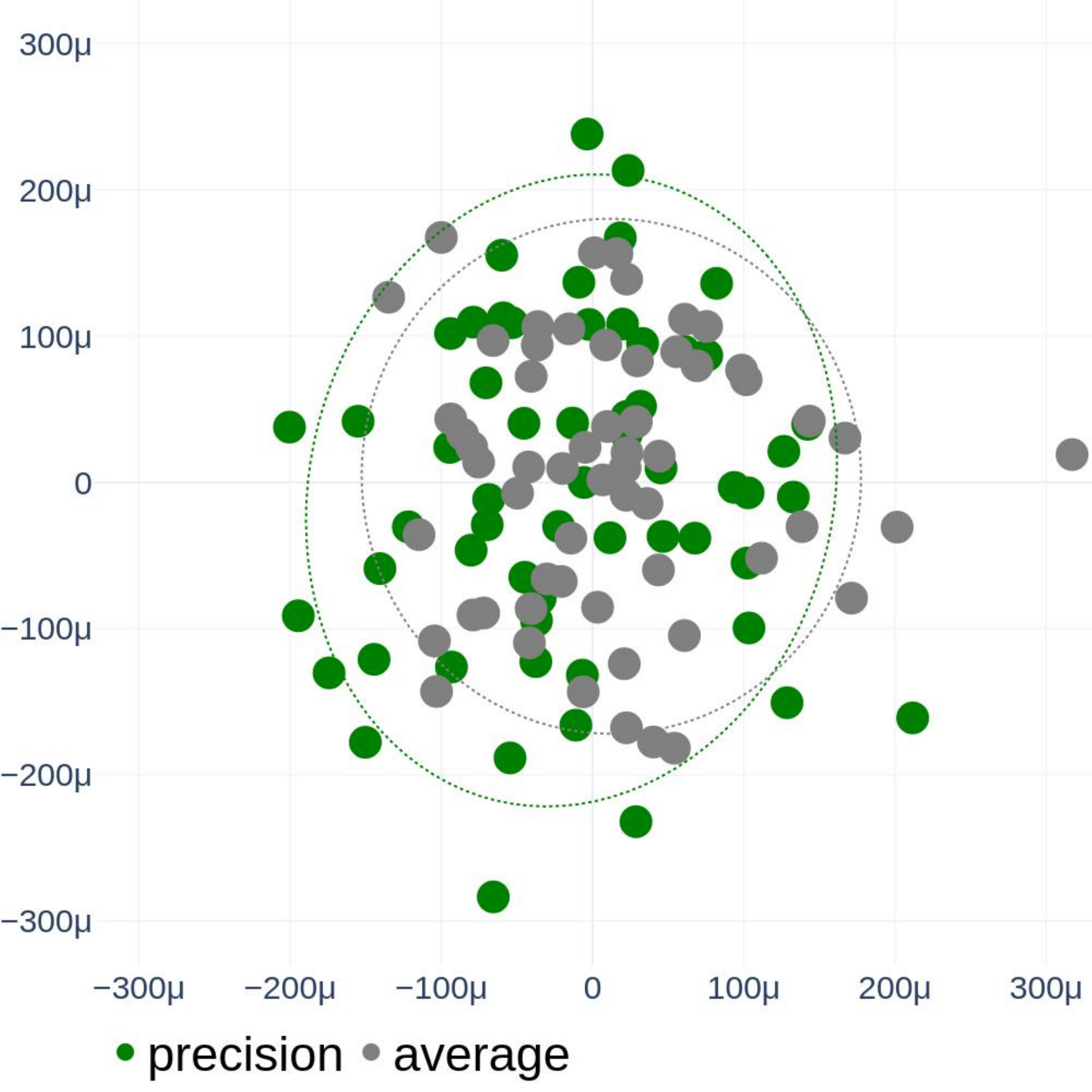}&
   \includegraphics[width=.33\linewidth]{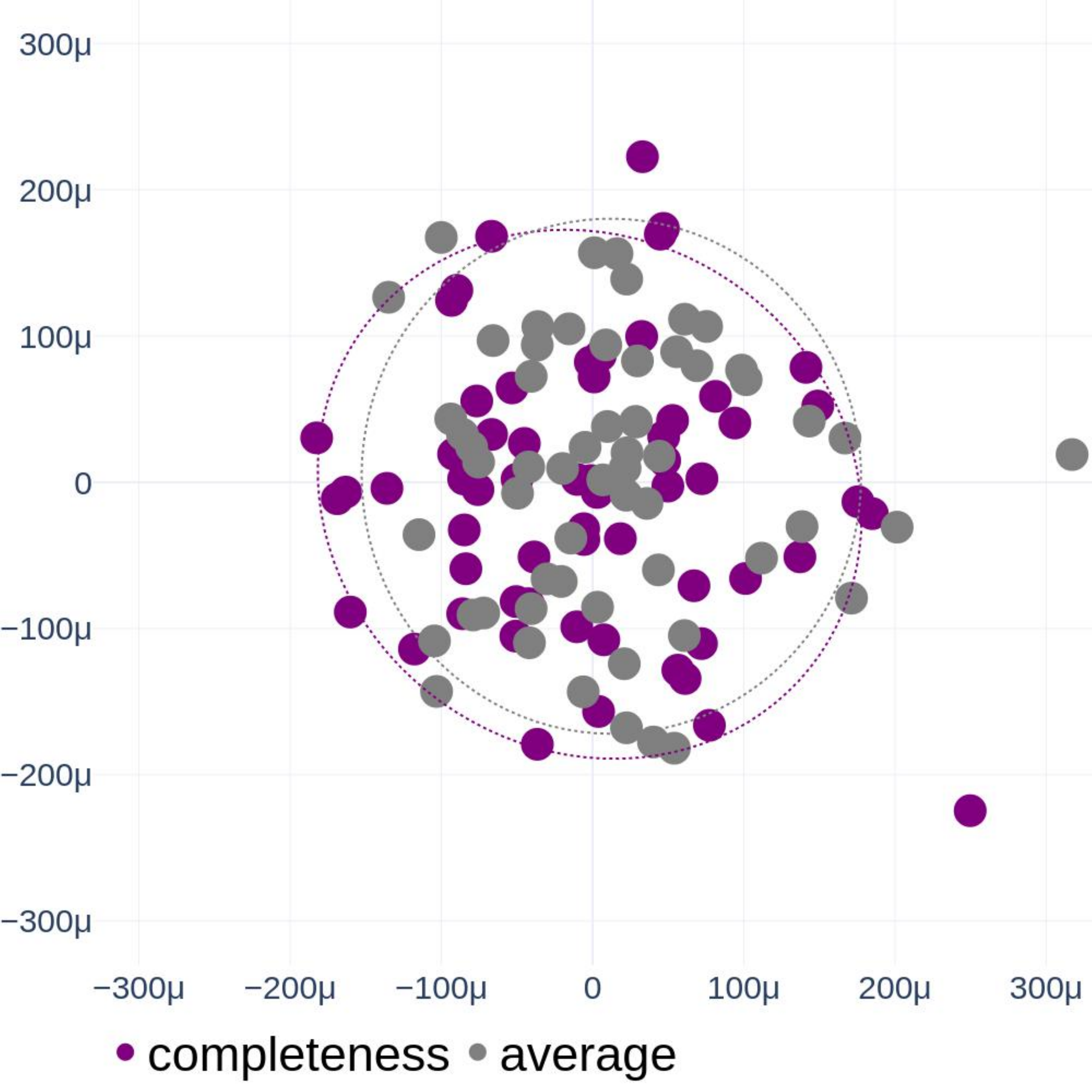}\\
   \includegraphics[width=.33\linewidth]{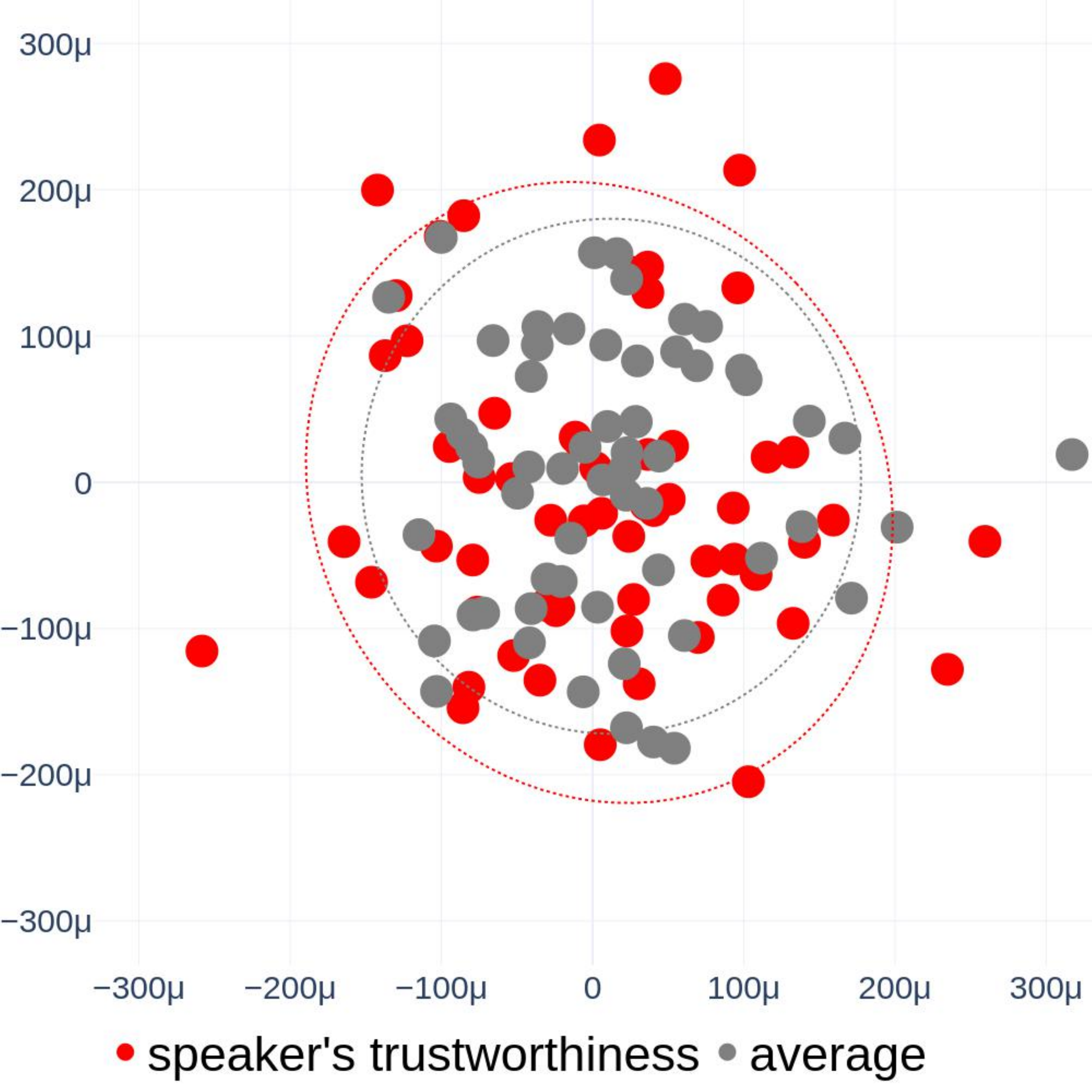}&
   \includegraphics[width=.33\linewidth]{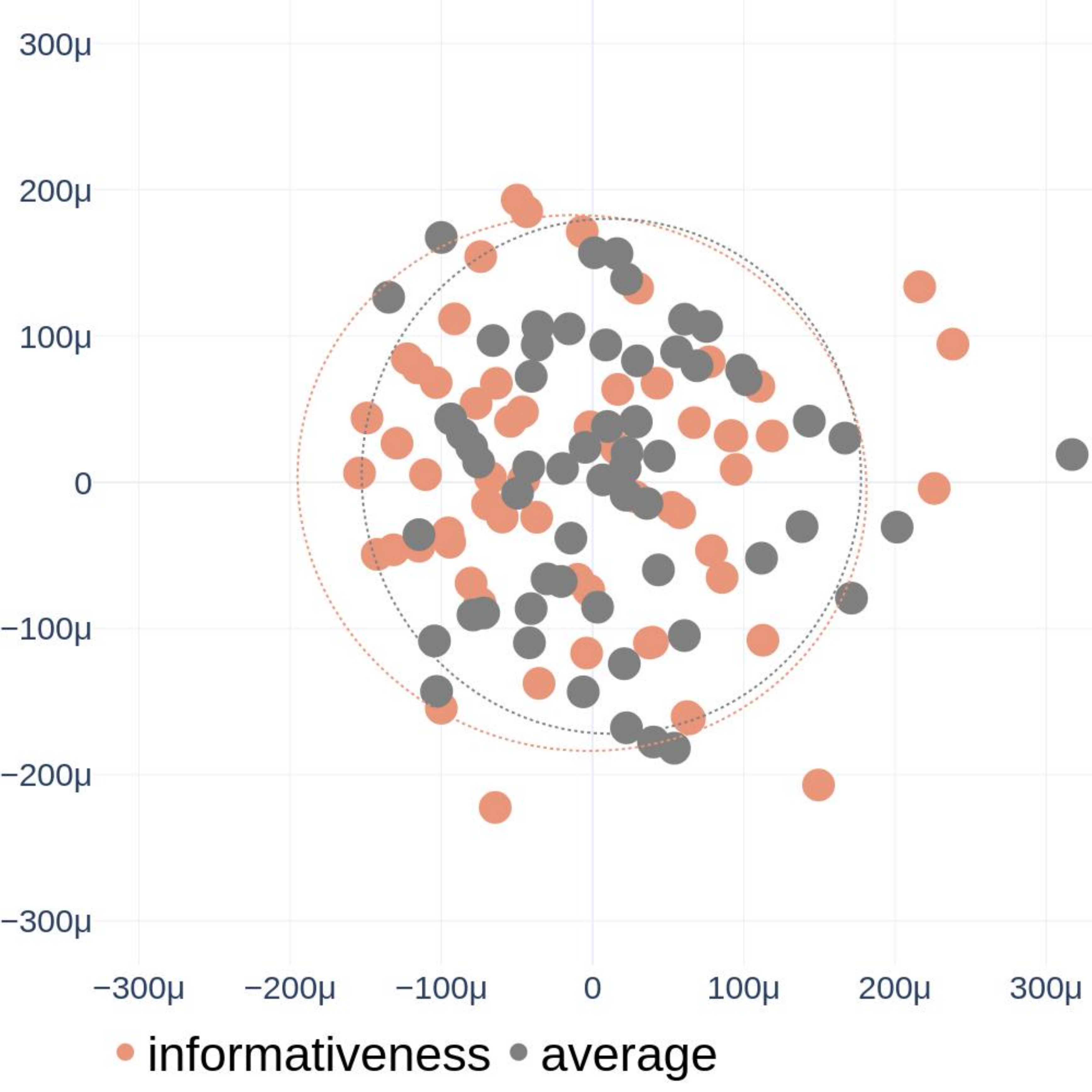}&
   \includegraphics[width=.33\linewidth]{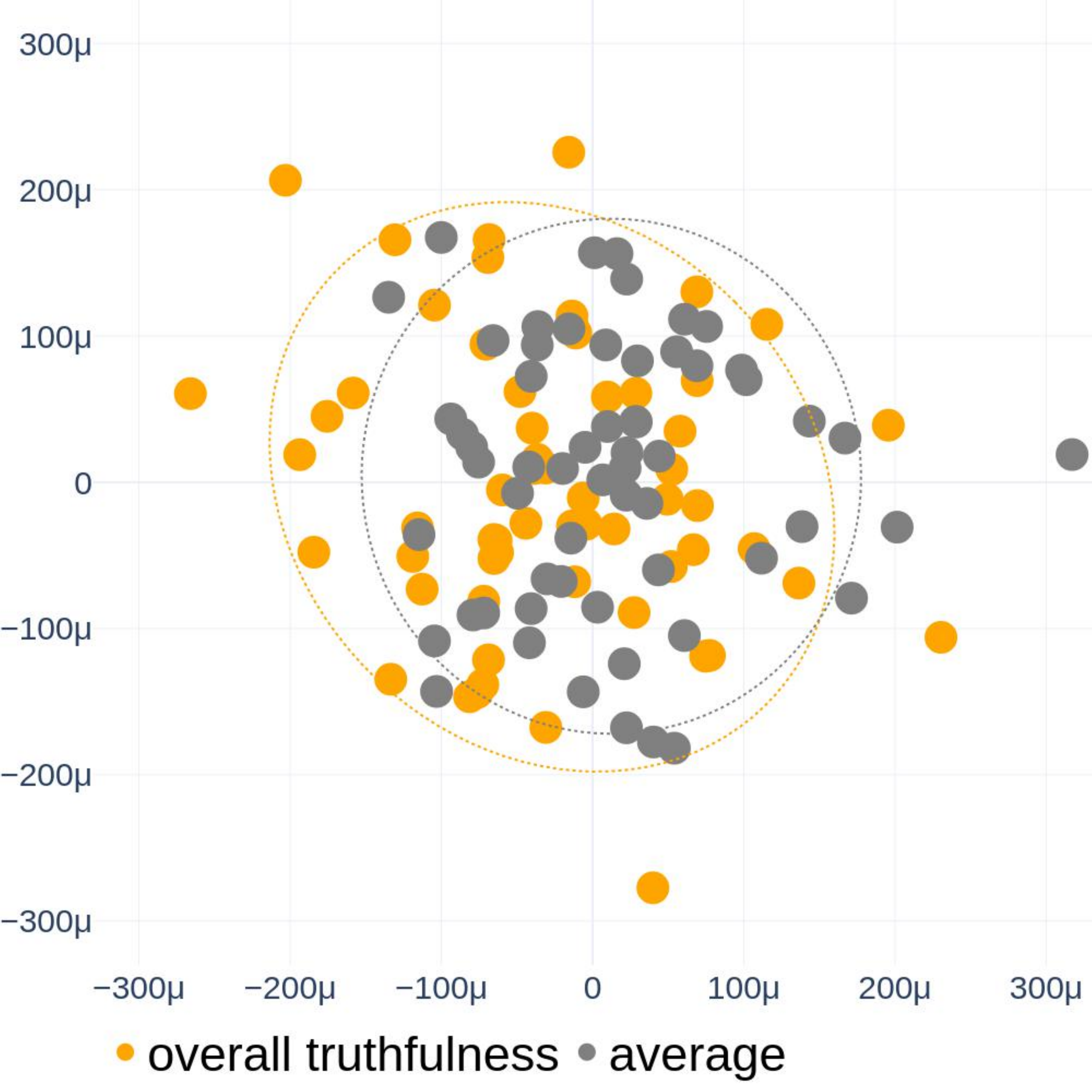}\\   
   \end{tabular}
 \caption{Visualization of the embeddings space. From top left: All dimensions together in the first plot, and then Correctness, Neutrality, Comprehensibility, Precision, Completeness, Speaker's Trustworthiness, Informativeness, and Overall Truthfulness, all of them compared to average. The coloring in the first visualization follows the legend used in the other plots. In each plot, we can observe how the distributions significantly overlap, thus indicating that the information loss due to the use of the average is rather limited.
}
   \label{fig:embeddings}
 \end{figure}

Averaging the embeddings introduces some information loss, but this loss is quite limited because the embeddings belong to the same semantic space. To investigate this aspect further, we show in Figure~\ref{fig:embeddings} the plots of the embeddings of each dimension and we compare them to the average embedding. These plots are obtained by using t-Distributed Stochastic Neighbor Embedding (t-SNE) to produce a meaningful bidimensional representation of the embeddings. Each plot includes an ellipsis representing the 95\% confidence interval for each of the sets of embeddings. Each set of embeddings can be thought of as a sample of the population of judgments that we can collect about the quality of the statements analyzed, weighted on the embedding representing the quality dimension's name. The significant overlap between the distribution of each set of embeddings and their average shows that the information loss is limited.

\textit{Linear Regression.}
We finally tested Linear Regression as a supervised approach based on weighted average word embeddings.
For each statement, we build an average word embedding of the assessments as mentioned in the previous approach. We calculate the word embedding of the corresponding ground truths, and we build a linear regression model that links the two. 
We run a 3-fold cross-fold evaluation, and every time our linear regression model predicts a verdict, we lookup in the embedding dictionary the closest term. The resulting average distance between the predicted judgment and the ground truth is 3.38 and the average sentiment difference is 0.41.
This methods improves the performance of the random selection baseline.
This indicates that the link between worker assessments and expert judgment is not straightforward as the previous approach hypothesized, but a linear model is already capable of capturing it to some extent. In the future, we plan to test more sophisticated models and to take workers profiles into consideration.

\section{Discussion and Conclusions}
\label{sect:discussion}

\subsection{Summary}
\label{sect:summary}

This work presents a study of the impact of crowdsourcing truthfulness judgments using multiple dimensions rather than just one. This allows for increased explainability of the collected labels as well as additional opportunities for quality control as crowd workers are asked to provide more input which can be cross-correlated. Our key findings are summarized below.

  \textit{Key Finding 1.} Addressing \ref{i:crowd-reliability}, we have provided extensive evidence that the truthfulness judgments provided by crowd workers over the seven dimensions of truthfulness are sound and reliable. The analyses of the internal agreement among workers do not show any issue with any of the dimensions. The agreement with the ground truth provided by experts is good when the same notion (i.e., Overall Truthfulness) is measured, and reasonable for the individual dimensions, with differences that can be justified by the meaning of each dimension. 

\textit{Key Finding 2.} Several analyses show that the seven dimensions are independent, not redundant, and measure different aspects (\ref{i:rrq2}). 
We have not been able to exploit this independence to combine the assessments on the single dimensions to obtain a higher agreement with the ground truth. 

\textit{Key Finding 3.} We found that different crowd workers behave differently; nevertheless, we have not been able to leverage such behavioral signals to improve the correlation between the aggregated crowd judgments and the ground truth of expert judgments
(\ref{i:worker-behavior}).  

\textit{Key Finding 4.} The  analyses on the informativeness of the different dimensions (\ref{i:dim-informativeness}) show that the crowd data are not easy to be generated automatically and that the different dimensions can be useful to understand the reasons behind the crowd worker's judgment.

\textit{Key Finding 5.} Finally, concerning \ref{i:machine-learning}, we have shown that signals derived from workers, and in particular their judgments, and search sessions, can be leveraged to effectively predict the expert verdicts, both for \politifact and \abc.

\subsection{Practical Implications}

From our analysis we can derive the following remarks, which are of practical use for researchers and practitioners who want to collect truthfulness using a multidimensional scale by means of crowdsourcing. 
\begin{itemize}
    \item Truthfulness judgments provided by crowd workers are reliable; workers are able to assess the truthfulness of political statements using a multidimensional scale.
    
    \item The agreement between the crowd judgments and the expert labels on the truthfulness dimension can be improved by aggregating the ground truth labels.
    
    \item 
    Researchers should expect that the workers values will tend to be skewed  towards the positive values of the Likert scale (i.e., \agree and \completelyagree); such behavior is present but less evident when the individual judgments are aggregated using the mean function.
    
    \item The seven dimensions we considered are independent; thus, researchers can re-use the same set of dimensions proposed in this work to collect truthfulness labels using crowdsourcing.
   
    \item Researchers should expect faster response times as the worker proceeds into the task since he/she will learn while doing it. For this reason, the same statement should be presented in different positions in the task to avoid any possible source of bias.
    
    \item Researchers should implement quality checks to obtain a high quality of the collected data.
    
    \item Researchers can use the gathered crowd labels to predict the expert judgments in a supervised learning scenario. In such a case, researchers should use the Random Forest algorithm as it provides the best effectiveness metrics, which are higher than baselines. 
    
    \item Researchers should expect no correlation between the judgments gathered on each single dimension and the corresponding set of computational measures that can be automatically computed for the same dimension; crowd workers and computational measures provide a different signal.
    
    \item Researchers should avoid to leverage internal agreement, as it appears to be rather low for all the dimensions and thus it does not provide any useful signal.
    
    \item Researchers should avoid the usage of naive techniques to combine dimensions as they do not improve the external agreement with expert labels.
    
\end{itemize}%

\subsection{Limitations}

A limitation of this work is that no ground truth for each of the seven dimensions exists. Having such information, we could perform a more direct comparison between the expert and worker annotations. However, we believe that such matter would constitute a different research project, complementary to this work, and not free from obstacles as information quality dimensions are more (e.g., comprehensibility) or less (e.g., accuracy) subjective. Furthermore, comparisons with an expert-provided ground truth might even be misleading since differences in the single dimensions may be due to subjectivity.
As future work, we plan to analyze expert annotated data once they are collected on all the dimensions, to check if there will be a difference with respect to this work.

Another limitation of this work consists in the aggregation method used to compare expert and worker labels, which might be questioned.
As reported in Section~\ref{subsect:task}, workers were asked to provide answers using the 5 level Likert scale \completelydisagree, \disagree, \neitheraord, \agree, \completelyagree. 
The adopted scale is clearly not nominal, since the considered categories are ordered. Furthermore, the considered scale does not represent a mere ranking, since the categories have a clear semantic meaning. Also, the perceived distances between the considered categories might be not consistent for all the workers (e.g., the perceived distance between
\completelydisagree and \neitheraord
might not be the double as the perceived distance between 
\completelydisagree and \disagree%
). %
On these bases, using the mean as an aggregation function might indeed be incorrect, since it assumes equidistant categories, and that might not be true for all the workers. Nevertheless, we have assumed that the adjacent categories are perceived as equidistant, also due to the labels that we used and that include a numeric value; if this assumption is correct, using the mean as an aggregation function should not introduce any error.
Furthermore, the alternatives are not free from limitations: both the median and the majority vote aggregation functions would discard possibly useful and significant signals and information. For a further discussion on label aggregation using the mean function when crowdsourcing truthfulness, see \citet[Section~3.3]{roitero2020crowd}. 

A further limitation of this study concerns the combination of dimensions to improve agreement between workers and expert labels.
As stated in Section~\ref{sect:truthfulness-dimensions}, we did try to combine the individual dimensions in a way that improves agreement between the crowd and expert judgments, and we also performed further analyses, but we could not find any increase in either the internal agreement among assessors or the external agreement between the crowd workers and experts. We plan to address this limitation in the future by considering more sophisticated techniques (as discussed in Section~\ref{sect:future-work}).

Another issue which is worth discussing, as it constitutes a possible limitation of this work, is the behavior of malicious workers. %
As detailed in Section~\ref{subsect:task}, we implemented a set of quality checks to ensure the high quality of the collected data. The workers which did not pass the quality checks (i.e., those are the malicious or non-diligent) were not allowed to submit the task, while we are pretty confident that workers which passed the quality checks performed the task in good faith. To verify this, we ran a set of analyses of the collected data (distribution of answers, time spent, behavioral analysis, etc.) and we did not observe any worker with a suspicious or outlier behavior. Thus, we are quite confident that workers which submitted the task are of high quality. Furthermore, we remark that the abandonment rate monotonically decreases as the worker continues to go through the task: most workers abandon the task right after reading the instructions, followed by those performing one judgment, etc. This is an evidence showing that a worker prefers to abandon the task if s/he finds it not appealing, rather than after having attempted to do it maliciously. Summarizing, to the best of our knowledge we have no reason to suspect any anomalous pattern in the collected data.

Also, a possible limitation of this work consists in the number and sample of statements  chosen for performing the analysis. We inspected the statements sampled by \citet{roitero2020crowd} and we did not notice any visible bias or difference with respect to the language level, terminology, length, etc. In future work we plan to address this limitation by performing a large scale analysis using more statements (as discussed in Section~\ref{sect:future-work}).

Finally, going back to the practical implications discussed in the previous subsection, we remark that although our findings show that multidimensional assessment is feasible and can be used to collect truthfulness labels in a crowdsourcing setting, we believe that we cannot claim that we have already reached the final long term goal of building a system to directly assess the truthfulness of statements as they appear on some social media using crowdsourcing \cite{demartini2020human}; this is an obvious future research direction, discussed among others in the following.%

\subsection{Future Work}\label{sect:future-work}

Overall, our experimental study allowed us to gather a large amount of data that will likely undergo additional analyses and applications (e.g., confidence values, selected URLs, complex combinations of the dimensions, text justifications, etc.) by the research community. A possible example of these future analyses could be the exploitation of the URLs collected  as evidence for each assessment and of the content of these web pages, to build a corpus of documents for each truthfulness level.
Another possible future extension of this work is to make use of the methodology detailed by \citet{sethi2019fact} to study the emotional aspects of misinformation as perceived by the crowd workers. We also plan to perform a large-scale experiment with a larger set of statements.

The seven dimensions that we used do not show correlation, with the only exception of Overall Truthfulness and Correctness. This means that such a set of dimensions can be re-used to collect truthfulness labels using crowdsourcing. Conversely, this also means that such a set may not be the optimal one. Therefore, another possible future extension of this work is to leverage the relationships and correlations between dimensions to characterize and find an optimal and definitive subset of dimensions to be used in future experiments.
In this work we employ an unsupervised approach to predict truthfulness judgments by computing static word embeddings and we leverage them to identify the semantic similarity between labels. Since such a method might suffers from information loss due averaging, in the future we plan to use different and more sophisticated approaches to compute word embeddings.%

We also remark that, when compared to systems or data collection approaches based on a single quality dimension, our approach favors explainability. Indeed, the assessment over multiple dimensions could allow to understand which facet(s) of the statement causes uncertainty and/or disagreement, and thus help to make an informed decision about the final truthfulness label of the statement. Hence, we believe it represents a step towards the design and development of systems to overcome the spreading of online misinformation that are robust, trustworthy, explainable, and transparent---which are aligned with the key principles that fact-checking organizations must follow.\footnote{\url{https://www.ifcncodeofprinciples.poynter.org/know-more/what-it-takes-to-be-a-signatory}} We plan to pursue this line of research in future work.%

\section*{Acknowledgments}

This work is partially supported by 
a Facebook Research award,
by the Australian Research Council
(DP190102141, DE200100064), by a MISTI - MIT International Science and Technology Initiatives - Seed Fund (MIT-FVG Seed Fund Project)
and 
by the project HEaD – Higher Education and Development - 1619942002 / 1420AFPLO1 (Region Friuli – Venezia Giulia).
We thank the reviewers for their comments; they provided insightful remarks that helped us to improve the overall quality of the paper.

\bibliography{mybibfile}

\newpage
\appendix
\onecolumn

\section{Task Instructions}\label{app:instructions}

This appendix contains of the instruction's text provided to each worker before starting the task.

\begin{tcolorbox}[breakable,title=Instructions provided to the workers]
In this task, you will be asked to assess the truthfulness of eight statements by means of seven specific quality dimensions.

First, you will be asked to fill in one questionnaire and to answer three questions. Then, we will show you \emph{8 statements} made by popular people (for example, political figures) together with the information of who made the statement and on which date. For each statement, we ask you to search for evidence using our custom search engine and to tell \emph{how much do you agree with considering the statement true in general (as opposed to false)}; that is, its overall truthfulness. We also ask you to mark the evidence found in terms of an URL as well as your self-confidence about the topic, i.e., if you consider \emph{yourself expert / knowledgeable about its topic (as opposed to novice/beginner)}.

Then, we ask you to assess seven specific \emph{quality dimensions} by stating your level of agreement with them. All your answers are given on a 5 level scale, i.e., they must  be selected among 5 different labels: (-2) Completely Disagree, (-1) Disagree, (0) Neither Agree Nor Disagree, (+1) Agree, (+2) Completely Agree. Each quality dimension is detailed in the following list. We provide a sample statement for each dimension so you can familiarize yourself with the seven dimensions. Please, note that there are some “positive” examples, (i.e., statements that completely agree with the current dimension), and “negative” examples, (i.e., statements that completely disagree with the current dimension). (Keep in mind that the examples are illustrative only, and it is likely that you may also need to use the rest of the labels in your answers). The seven dimensions we consider are the following:

\begin{itemize}
    \item \emph{Correctness}: the statement is expressed in an accurate way, as opposed to being incorrect and/or reporting  mistaken information
    \begin{itemize}
        \item Example (which label is: +2 Completely Agree): “It’s illegal to treat a minor without parental consent in the U.S. Even as hospitals are limiting visitors, minors will always be allowed to have one guardian present.”
    \end{itemize}
    \item \emph{Neutrality}: the statement is expressed in a neutral / objective way, as opposed to subjective / biased
    \begin{itemize}
        \item Example(which label is: -2 Completely Disagree): “The Labor Party has repeatedly claimed the Coalition needs to make cuts of \$70 billion to vital services to balance the budget.”
    \end{itemize}
    \item \emph{Comprehensibility}: the statement is comprehensible / understandable / readable as opposed to difficult to understand
    \begin{itemize}
        \item Example (which label is: +2 Completely Agree) “Florida ranks first among the nations for access to free prekindergarten."
    \end{itemize}
    \item \emph{Precision}: the information provided in the statement is precise / specific, as opposed to vague
    \begin{itemize}
        \item Example (which label is: -2 Completely Disagree): “There were more deaths after the gun bans from guns than there were in the three years before Port Arthur”
    \end{itemize}
    \item \emph{Completeness}: the information reported in the statement is complete as opposed to telling only a part of the story
    \begin{itemize}
        \item Example (which label is: -2 Completely Disagree): “We inherited a broken test for COVID-19.”
    \end{itemize}
    \item \emph{Speaker's trustworthiness}: the speaker is generally trustworthy / reliable as opposed to untrustworthy / unreliable / malicious
    \begin{itemize}
        \item Example (which label is: -2 Completely Disagree): “Says video shows “the Chinese are destroying the 5G poles as they are aware that it is the thing triggering the corona symptoms.”
    \end{itemize}
    \item \emph{Informativeness}: the statement allows us to derive useful information as opposed to simply stating well known facts and/or tautologies.
    \begin{itemize}
        \item Example (which label is: +2 Completely Agree): “2019 coronavirus can live for up to 3 hours in the air, up to 4 hours on copper, up to 24 hours on cardboard up to 3 days on plastic and stainless steel.”
    \end{itemize}
\end{itemize}

If you wish to change a previously given judgment, you can use the Back and Next buttons to navigate the task and revisit your answers. Please note that the statements are not presented in any particular order. You might see many good statements, many bad ones, or any combination. Try not to anticipate, and simply rate each statement after reading it. Note that you’ll need to answer \emph{all questions and fill in every field} in order to proceed in the task, otherwise you will not be able to proceed to the following steps. Note that there are some quality checks throughout the task, and if you do not perform these correctly you will not be able to terminate the task and get paid. The data from this task is being gathered for research purposes only. Participation is entirely voluntary, and you are free to leave the task at any point.
\end{tcolorbox}
\end{document}